\documentclass{elsart}
\usepackage{color,setspace,times}
\usepackage{amssymb,amsmath,graphicx,color,rotating,subfigure,url}
\usepackage[square,sort&compress,comma]{natbib}
\usepackage{lineno}

\journal{Physica A}

\begin{document}

\begin{frontmatter}

\title{Trading Model with Pair Pattern Strategies}


\author[SB,CES]{\small{F. Ren\thanksref{Addr}}},
\author[fru,pdru]{\small{Y. C. Zhang}}
\address[SB]{School of Business, East China University of Science and Technology, Shanghai 200237, PR China}
\address[CES]{Center for Econophysics Studies, East China University of Science and Technology, Shanghai 200237, PR China}
\address[fru]{Departement de Physique, Universit\'{e} de Fribourg, Perolles CH-1700, Switzerland}
\address[pdru]{Physics Department, Renmin University, Beijing, PR China}
\thanks[Addr]{Corresponding author. {\it E-mail address:}\/
fren@ecust.edu.cn}

\begin{abstract}
A simple trading model based on pair pattern strategy space with
holding periods is proposed. Power-law behaviors are observed for
the return variance $\sigma^2$, the price impact $H$ and the
predictability $K$ for both models with linear and square root
impact functions. The sum of the traders' wealth displays a positive
value for the model with square root price impact function, and a
qualitative explanation is given based on the observation of the
conditional excess demand $\langle A|u \rangle$. An evolutionary
trading model is further proposed, and the elimination mechanism
effectively changes the behavior of the traders highly performed in
the model without evolution. The trading model with other types of
traders, e.g., traders with the MG's strategies and producers, are
also carefully studied.
\end{abstract}


\end{frontmatter}

\section{Introduction}

The standard Minority Game (MG) introduced and studied by Challet
and Zhang \cite{cha97,zha98} was initially designed as a
simplification of Arthur's famous El Farol's Bar problem
\cite{art94}. It describes a system in which many heterogeneous
traders adaptively compete for a scarce resource, and it captures
some key features of a generic market mechanism and the basic
interaction between the traders and public information. However, it
is a highly simplified model, not suitable to compare with real
financial market trading. To make it more realistic in comparison
with the real markets, different variations
\cite{cha01,jef00,cha01a,gia02,and02,MGw04,ren06} of the standard MG
are consequently proposed. For example, the inactive strategy is
introduced, which grants the traders with the possibility of not
trading and thus the number of traders actively trading at each time
step varies throughout the game. This type of extension is called
the grand canonical MG, and it produces the main characteristics of
the stock markets, e.g., the "fat tail" return distribution and the
long-range volatility correlations.

In the standard MG and most of its variations, the strategies give
predictions for the next time step based on the current state of the
history, and upon which traders make instantaneous trading action
with an horizon not more than one time step. For example, the
MG-based models with dynamic capitals \cite{jef00,cha01a}, in which
the wealth of the traders is updated according to the current
trading price and has no relation with the trades they have made
previously. The payoff raised from the price change at the next time
step is grant to the trader immediately after a single trade. In
fact traders have asset holding periods and make profits from price
difference between two consecutive trading actions of buying and
selling in the real stock markets. To find a strategy space with
holding periods, that traders open or close their positions and hold
their positions reasonably is a great challenge for modeling
speculation.

Recently, a new model based on a simple pattern-based strategy space
with holding periods is proposed by Challet \cite{cha08}. In this
model, the strategy space is composed of a sequence of patterns,
i.e., history signals. Traders open or close positions when the
current pattern is the pattern listed in the strategy space, and
hold positions between patterns. The kind of position he/she might
take, i.e., buying or selling is determined by the average price
return between two consecutive occurrences of patterns. The
explicitly trading action of buying or selling is not fixed with the
patterns. A simplest case that the strategy space consists of only
one pair of patterns is mainly considered.

Inspired from Challet's work, we introduce a new pattern-based
speculation model in which the patterns strategy space is split into
several sub-spaces composed by pairs of patterns. Different from
Challet's model, we defined the explicitly trading actions of the
pair patterns. One strategy consists of a pair of patterns, which
denote the history signals for buying and selling. It is reasonable
to assume that the trader base their decisions on patterns or
history signals since they may have some experiences and know when
to buy or sell according to the history signals. The order of the
pair patterns does not make any sense, which means the position can
be opened by buying/selling if the history signal for buying/selling
comes first. Therefore, the trader should buy first if he/she wants
to sell, or buy only after he/she sells in this model. That is
exactly the case in the real stock markets.

The MG is a negative sum game due to the minority essence of its
payoffs. The traders compete for the limited market resource and
only those traders in the minority group are rewarded. Challet's new
model keeps the minority-game payoffs impacted by his/her own trades
when the trader opens and close the position. Furthermore, it
introduces a new term of majority-game payoffs raised from the
contribution of other traders during his/her holding periods. The
sum of the payoffs therefore depends on the trading frequency and
reaction time of each trader, and thus makes the dynamics relatively
complicated.

For the real financial market, the sum of the social wealth should
be positive due to the general increase of the social productivity.
Traders are willing to trade in the market which has a positive sum
of the social wealth or at least has a zero sum. The purpose of this
paper is to construct a model with holding periods which has a zero
sum or positive sum. We first assume a zero wealth sum in our model
by introducing a simple mid-price dynamics with a linear price
impact function. Furthermore, a square root impact function revealed
by the empirical study of the real stock market \cite
{ple02,gab03,has91} is considered, and the model consequently tends
to be a positive sum game. The payoff for each trader is naturally
determined by the difference between the selling price and the
buying price. We name this model as trading model due to the trading
essence of the pair patterns.

In Sec. II, we first introduce a pattern-based speculation model by
Challet, and then introduce our trading model with pair pattern
strategies. A comprehensive comparison between the definition of the
strategy space and the payoffs of our model and that of Challet's
model is detailed presented. Some numerical results of our trading
model are subsequently presented. In Sec. III, a dynamic evolution
mechanism is introduced to the trading model, and the process how
the traders are washed out and how their wealth evolves are
carefully studied. In Sec. IV, other types of traders are introduced
to the trading model, and Sec. V contains the conclusion.

\section{Trading Model with Pattern-based Strategy Space}

\subsection{A pattern-based speculation model by Challet}

The model proposed by Challet \cite{cha08} consists of N traders,
and they base their decisions on patterns. Each trader $i$ is able
to recognize $S$ patterns $\mu_{i,1},\cdot \cdot \cdot,\mu_{i,S}$,
drawn randomly and uniformly from $\{1,\cdot \cdot \cdot,P\}$ at the
beginning of the game and kept fixed throughout the game. Each
trader $i$ keeps track of the cumulative price return between two
consecutive occurrences of patterns, denoted by
$U_{i,\mu\rightarrow\nu}$, where $\mu,\nu\in\mu_{i,1},\cdot \cdot
\cdot,\mu_{i,S}$ and $\mu\neq\nu$. At time $t$, the trader may wish
to open position only when the current history signal $\mu(t)$ is in
his pattern list, that is, $\mu(t)\in \{\mu_{i,1},\cdot \cdot
\cdot,\mu_{i,S}\}$. The kind of position he/she might take,
$a_{i}(t)=0,\pm1$ denoting inactive, buying and selling, is
determined by average price return between two consecutive
occurrences of patterns: if $|U_{i,\mu\rightarrow\nu}(t)|>\epsilon
t_{\mu\rightarrow\nu}$, where $t_{\mu\rightarrow\nu}$ is the total
number of time-steps between patterns $\mu$ and $\nu$ and
$\epsilon>0$ is a parameter, one buys a share
($U_{i,\mu\rightarrow\nu}(t)>0$) or sells a share
($U_{i,\mu\rightarrow\nu}(t)<0$), and then hold his/her position
until $\mu(t')=\nu$. The excess demand is $A(t)=\sum_{i=1}^{N}
a_i(t)$. A linear price impact function is considered, and thus the
price return is simply defined as
\begin{equation}
r(t)=p(t+1)-p(t)=A(t). \label{e10}
\end{equation}

Assume that $\mu(t_{\mu})=\mu$ and $t_{\nu}$ is the first subsequent
occurrence of pattern $\nu$, the cumulative price return
$U_{i,\mu\rightarrow\nu}$ between pattern $\mu$ and $\nu$ evolves
according to
\begin{eqnarray}
U_{i,\mu\rightarrow\nu}(t_{\nu}+1)=U_{i,\mu\rightarrow\nu}(t_{\mu})&+&p(t_{\nu}+1)-p(t_{\mu}+1)\nonumber\\
\nonumber\\
&-&(1-|a_{i}(t_{\mu})|)\zeta_{i}[A(t_{\nu})-A(t_{\mu})], \label{e20}
\end{eqnarray}
where $\zeta_{i}$ is a naivety factor indicating the reaction time
of trader $i$. By adjusting the value of the parameters $\epsilon$
and $\zeta_{i}$, we observe that the model exhibits a withdraw
phenomena in its price evolution.

If trader $i$ decides to open a position at time $t_{i,\mu}$ and
close his/her position at time $t_{i,\nu}$, then his/her payoff is
\begin{eqnarray}
a_i[p(t_{i,\nu}+\delta t_{i,\nu})-p(t_{i,\mu}+\delta t_{i,\mu})]=
&-& a_i A(t_{i,\mu},\delta t_{i,\mu})+ a_i \sum_{t_{i,\mu}+\delta
t_{i,\mu}<t\leq t_{i,\nu}} a(t) \nonumber\\
&-&(-a_i)A(t_{i,\nu},\delta t_{i,\nu}), \label{e30}
\end{eqnarray}
\begin{equation}
A(t_{i,\mu},\delta t_{i,\mu})=\sum_{t_{i,\mu}<t\leq t_{i,\mu}+\delta
t_{i,\mu}}a(t). \label{e40}
\end{equation}
$\delta t_{i,\mu}$ and $\delta t_{i,\nu}$ are one's reaction time
when he/she opens and closes position, which may due to
communication delays and the time needed to make a conscious
decision. The first and the last terms are minority-game payoffs,
which can be easily recognize by their '-' sign: the trader is
rewarded if he/she takes an action opposite to the majority of the
orders executed during the time delay. The central term which has a
'+' sign could be regarded as a delayed majority-game payoff: the
trader is rewarded if he/she takes an action consequently proved to
be consistent with the majority of the orders executed during the
holding period. Therefore, the traders' wealth depends on the
situation of the market: the relative importance of minority games
decreases as the trading frequency decreases and increases as the
reaction time of each trader increases.

\subsection{Trading model with pair pattern strategies}

The trading model takes the form of a repeated game with a certain
number of traders $N$. Different from challet's model, we split the
pattern strategy space into several sub-spaces in units of pair
patterns. A strategy consists of a pair of patterns or history
signals with explicit trading actions, labeled as $(\mu,\nu)$, where
$\mu$ is for buying and $\nu$ is for selling. A pattern or history
signal records the possible status of the $m$ most recent outcomes
of the price change. Since there are a total of $2^m$ possible
patterns and the patterns of one strategy should not repeat, there
are a total of $2^m*(2^m-1)$ probable pairs of patterns.

At the beginning of the game, each trader randomly picks $S$
strategies from the full strategy space and keep them fixed
throughout the game. Each trader $i$ keeps track of the cumulative
performance of his/her pair pattern strategy $s, s=1,...,S$ by
assigning a virtual score $U_{i,s}$ to it. The initial scores of the
strategies are set to be zero. At time $t$, each trader $i$ adopts
the strategy with the highest score $s_i(t)$, and checks if either
of the two patterns of the highest score strategy is consistent with
the history at that moment. If the pattern for buying occurs first,
the trader opens an position by buying and holds the position until
the pattern for selling occurs. Then the trader closes the position
by selling. The trader can also open the position by selling if the
pattern for seeling occurs first and then close the position by
buying. Therefore, the model is symmetric. The action will then be
$a_i(t)=0,\pm1$, denotes inactive, buying and selling. The excess
demand is defined as $A(t)=\sum\limits_{i=1}^{N} a_i(t)$.

We define a simple price dynamics of returns with a linear price
impact function the same as Eq. ({\ref {e10}}). Assuming that one of
the patterns of a strategy occurs at time $t_1$ and $t_2$ is the
first subsequent occurrence of the other pattern, the score of the
strategy is updated according to the price difference between these
two patterns as
\begin{equation}
U_{i,s} (t_2+1)=U_{i,s} (t_1)+ p(t_{s_i,\nu}+1)-p(t_{s_i,\mu}+1),
\label{e50}
\end{equation}
where $s_i$ is the strategy of trader $i$. If the pattern for buying
$\mu$ occurs first $t_{s_i,\mu}=t_1$ and $t_{s_i,\nu}=t_2$, and if
the pattern for selling $\nu$ occurs first $t_{s_i,\nu}=t_1$ and
$t_{s_i,\mu}=t_2$. Therefore, the payoff of the strategy is
determined by the profit made from the strategy if it is adopted. In
our model, we assume the traders are sophisticated and compute
perfectly the price return, and this makes Eq. ({\ref {e50}}) look
like Eq. ({\ref {e20}}) with $\zeta_{i}=0$.

A wealth $W_i$ is assigned to each trader $i$. If trader $i$ decides
to open a position at time $t_1$ and closes it at time $t_2$, the
wealth is updated according to the price return between these two
trades as
\begin{equation}
W_{i} (t_2+1)=W_{i} (t_1)+ p(t_{i,{\nu}}+1)-p(t_{i,{\mu}}+1).
\label{e60}
\end{equation}
If the trader opens a position by buying first $t_{i,{\mu}}=t_1$ and
$t_{i,{\nu}}=t_2$, and if the trader opens a position by selling
first $t_{i,{\nu}}=t_1$ and $t_{i,{\mu}}=t_2$.

However, the model with payoffs defined as above is a negative-sum
game, which means that the sum of the wealth of all the traders is
negative. Considering the simplest case that the market only has one
trader, the payoff of his/her wealth is always $-1$ no matter what
kind of position he/she might take first. We assume that the model
with linear price impact function has a nature of zero-sum. Inspired
by the works in Refs.\cite{and02,far98}, a middle price is
introduced $p(t'+1)=\frac{1}{2}(p(t+1)+p(t))$. Supposing that not
all the shares are executed at the price immediately after the
trades, the traders make trades at the middle price on average. The
price return defined with the linear price impact function is
\begin{equation}
r(t')=p(t'+1)-p(t') =\frac{1}{2}(A(t')+A(t'-1)), \label{e70}
\end{equation}
where $A(t')=A(t)$. The score of the strategy and the wealth of each
trader are consequently updated according to the middle price,
replacing $t$ by $t'$ in Eqs. ({\ref {e50}}) and ({\ref {e60}}). We
simply assume that the trader has no reaction time, and therefore
the payoff looks like Eq. ({\ref {e30}}) with $\delta t=0$. The sum
of the traders' wealth at time $t'$ is
\begin{equation}
\sum_{i} W_{i}(t')= \sum_{i}\sum_{t'_{i,1},t'_{i,2}}^{1\leq
t'_{i,1},t'_{i,2}\leq t'} [\frac{1}{2} A(t'_{i,1})+ \frac{1}{2}
A(t'_{i,2})+\sum_{t'_{i,1}< \tau <t'_{i,2}}A(\tau)], \label{e80}
\end{equation}
where $t'_{i,1}$ is the time trader $i$ opens a position and
$t'_{i,2}$ is the time trader $i$ closes it. For the model with line
price impact function, the sum of the first two terms approximately
equals zero since the number of positions opened by traders equals
to that closed by traders. The sum of the traders' wealth
consequently depends on the cumulative access demand contributed by
the traders during the holding periods.

Recent empirical study shows that the volume imbalance seems to have
a square root impact on the price return \cite {ple02,gab03,has91}.
Therefore, we also introduce a square root impact function to the
price return as
\begin{equation}
r(t')=p(t'+1)-p(t')
=\frac{1}{2}(sign(A(t'))\sqrt{|A(t')|}+sign(A(t'-1))\sqrt{|A(t'-1)|}).
\label{e90}
\end{equation}
The score of the strategy, the wealth of each trader and the sum of
the traders' wealth are consequently updated according to this price
dynamics.

\subsection{The results}

\begin{figure}[htb]
\centering
\includegraphics[width=6.5cm]{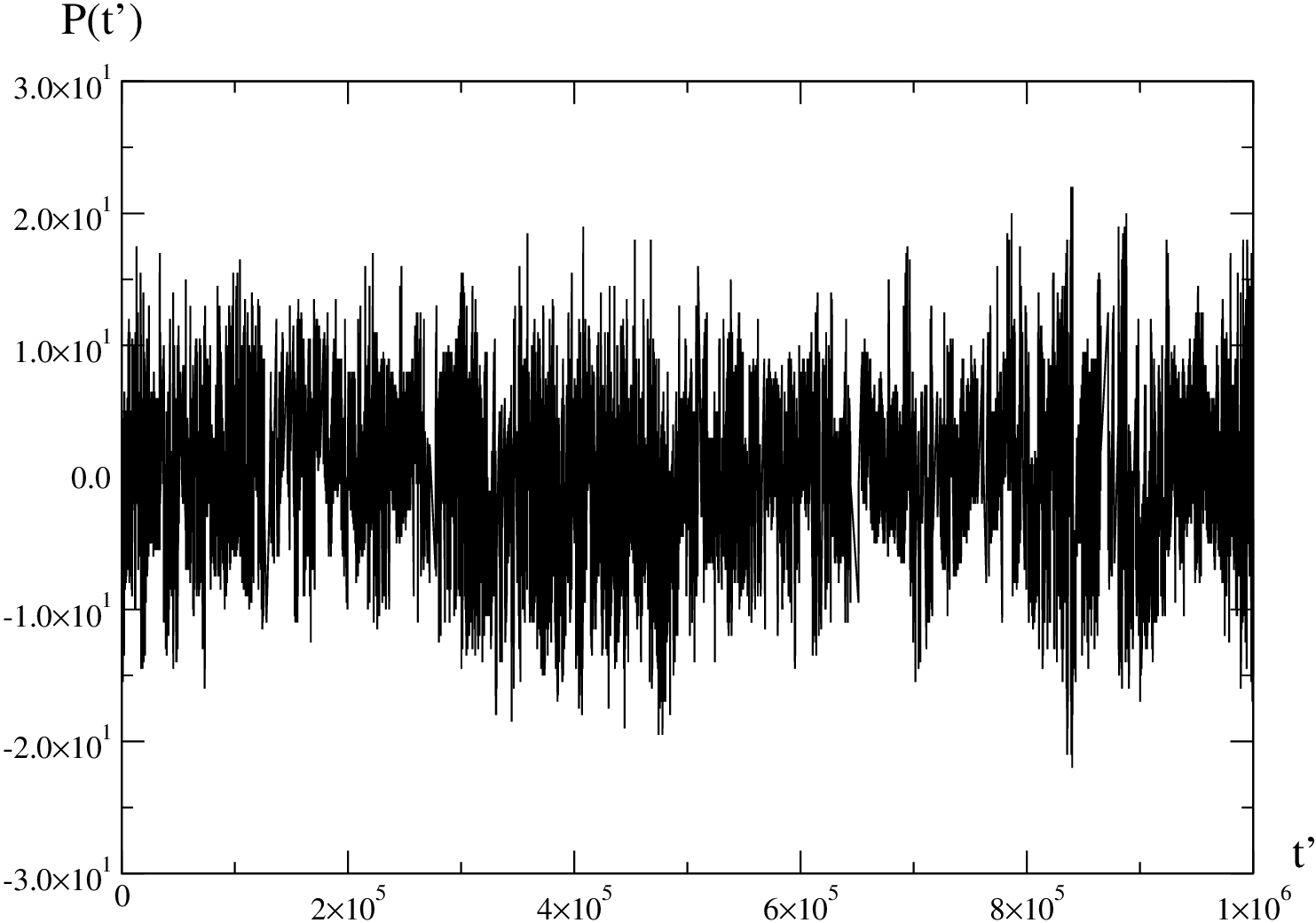}
\includegraphics[width=6.5cm]{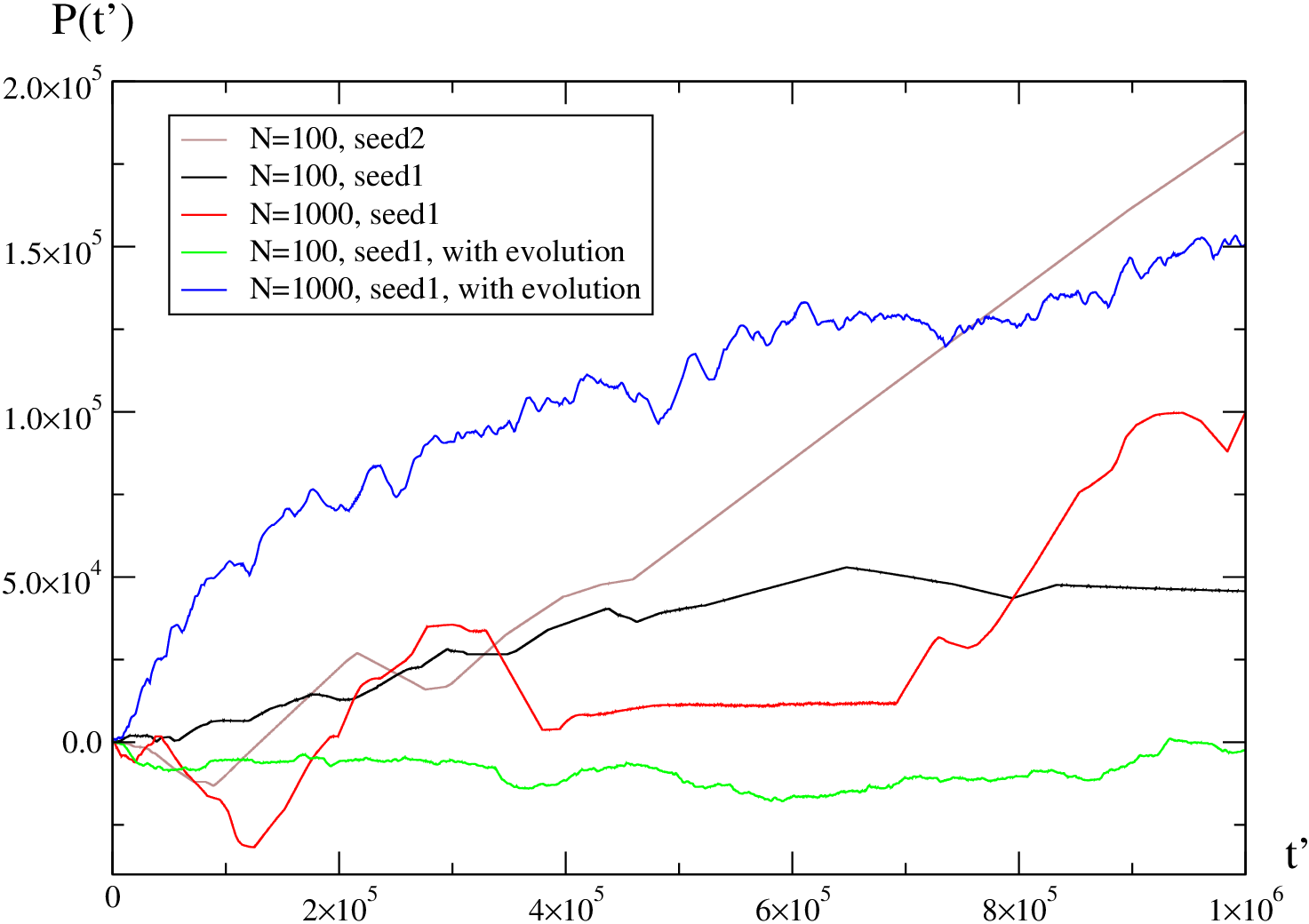}
\caption{\label{f1} (a) Price evolution of the trading model with
linear price impact function at $N=100$, $S=2$ and $m=3$. (b) Price
evolution of the trading model with square root price impact
function for different values of the parameter $N$ and different
initial seeds at $S=2$ and $m=3$. $10^{6}$ data are collected after
$500$ iterations for equilibrium.}
\end{figure}

Numerous numerical simulations are performed for this trading model
with pair pattern strategies, and the results for $S=2$ and $m=3$
are mainly reported. The price evolution of the trading model with
linear price impact function for $N=100$, and square root price
impact function for $N=100,1000$ and different initial seeds are
plotted in Fig.~\ref {f1}(a) and (b). For the model with linear
price impact function, the curve for $N=100$ fluctuates
symmetrically around zero, and the curve for other value of the
parameter $N$ behaves similar to that of $N=100$ (not shown in
figure). For the model with square root price impact function, it
seems that the price fluctuates similar to that of the financial
markets. However, we observe some attractors for some singular runs,
for example the curve for $N=100$ with $seed2$ changes suddenly
close to $t'=5.0*10^{5}$ and then the system stuck in a string of
periodical history states which may leads to the abnormal increase
of the price.

\begin{figure}[htb]
\centering
\includegraphics[width=4.5cm]{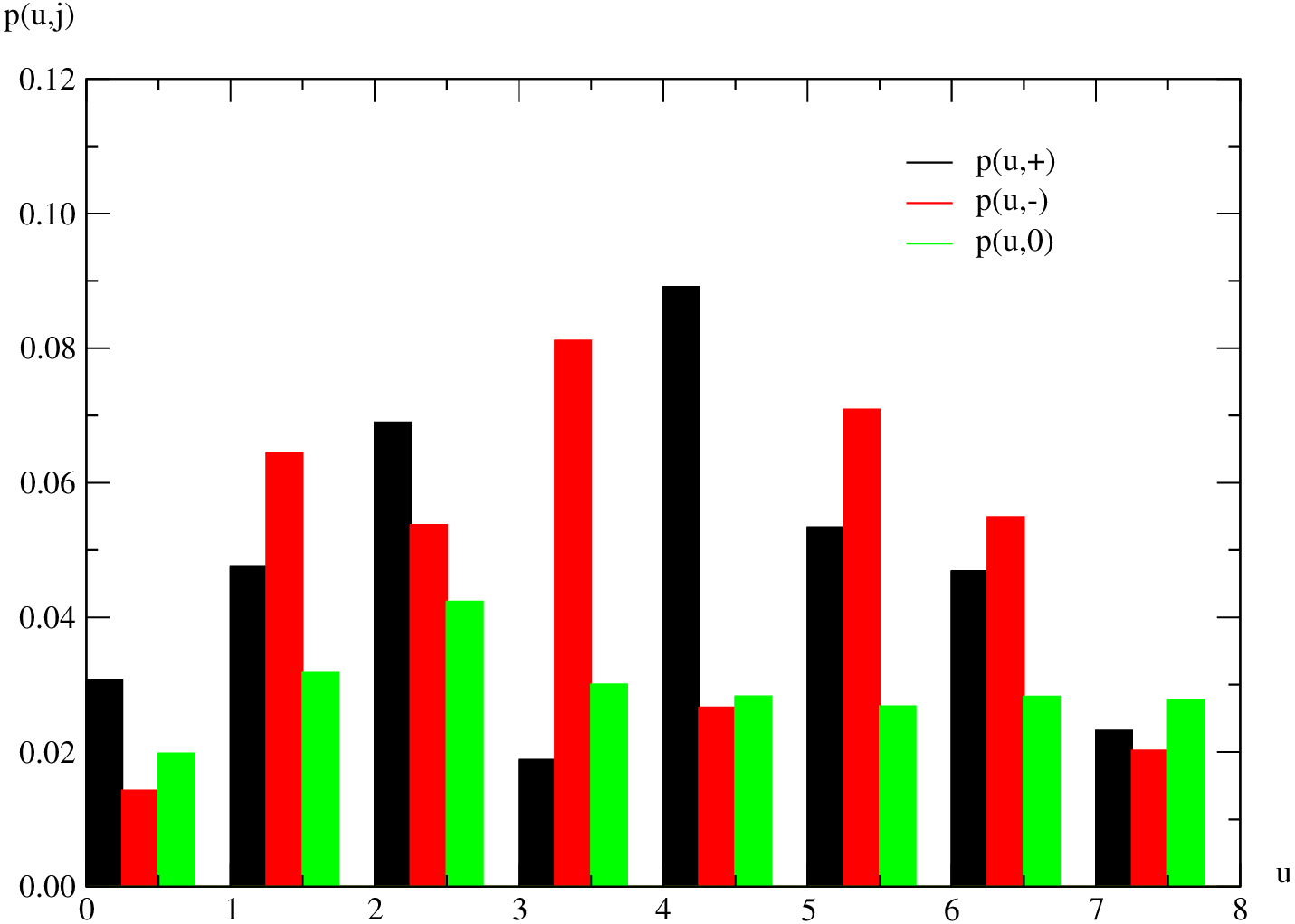}
\includegraphics[width=4.5cm]{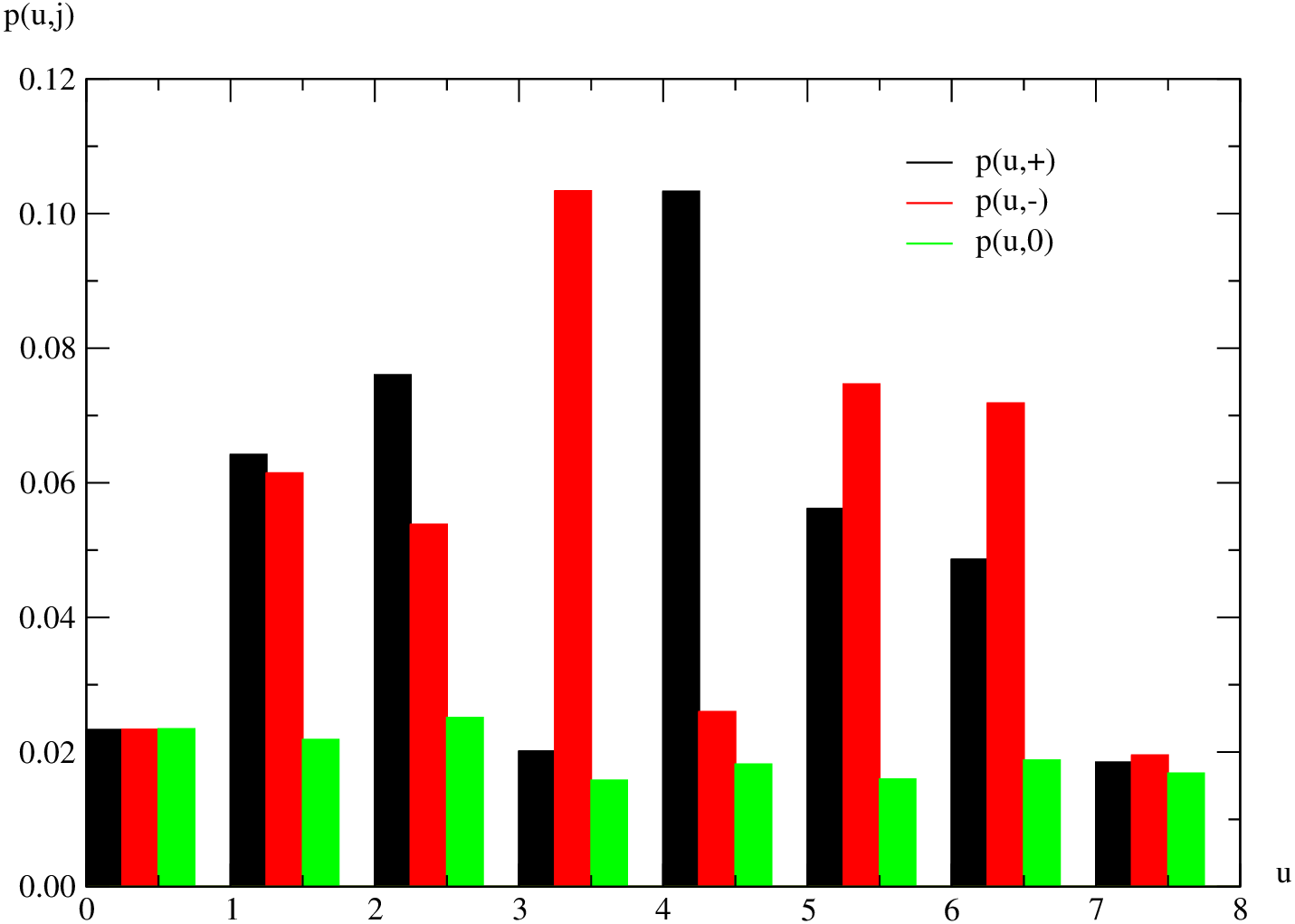}
\includegraphics[width=4.5cm]{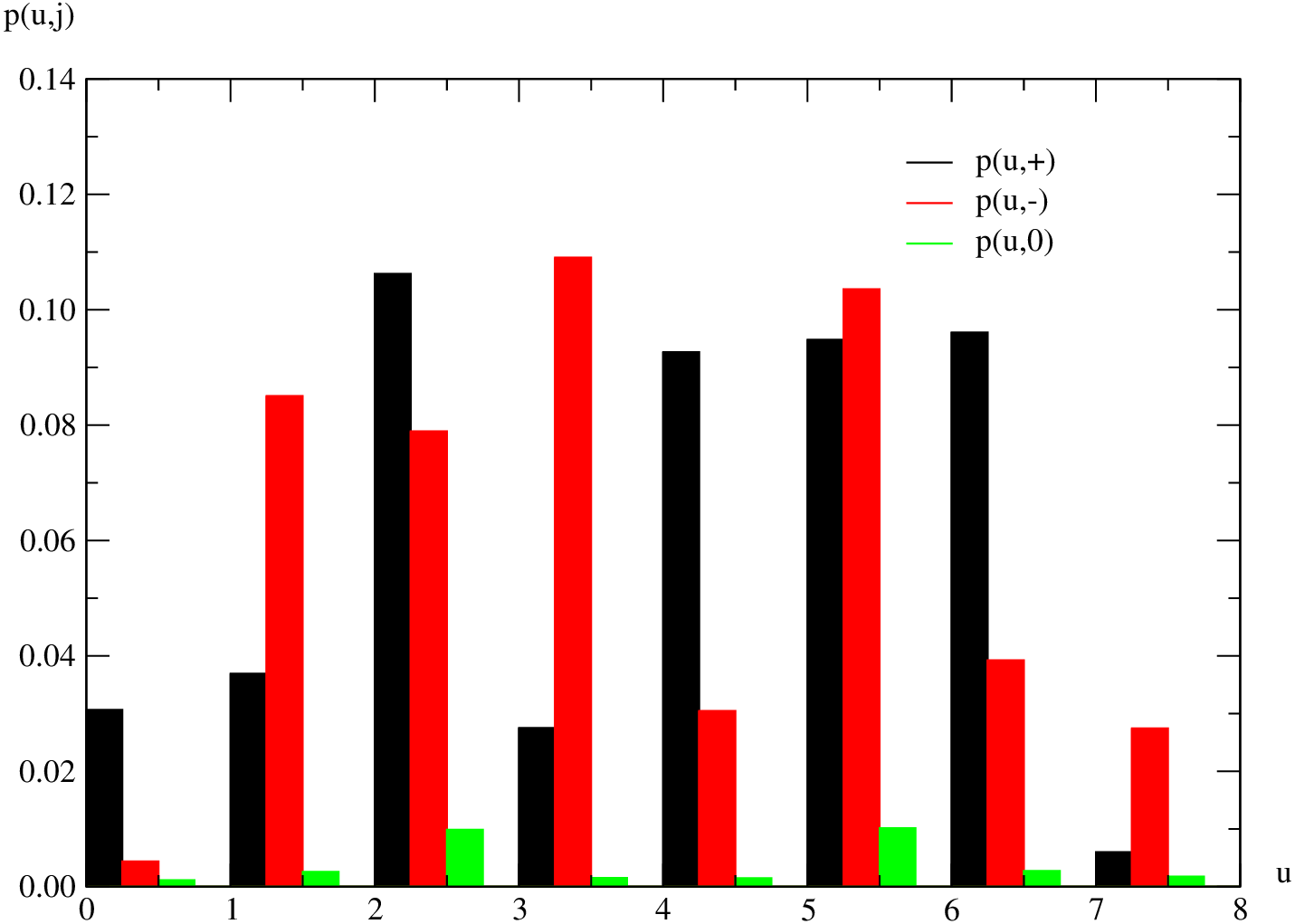}
\caption{\label{f2} Conditional probability $p(u,j)$ of the model
with linear price impact function for: (a) $N=50$, (b) $N=100$, (c)
$N=1000$ at $S=2$ and $m=3$.}
\end{figure}

\begin{figure}[htb]
\centering
\includegraphics[width=4.5cm]{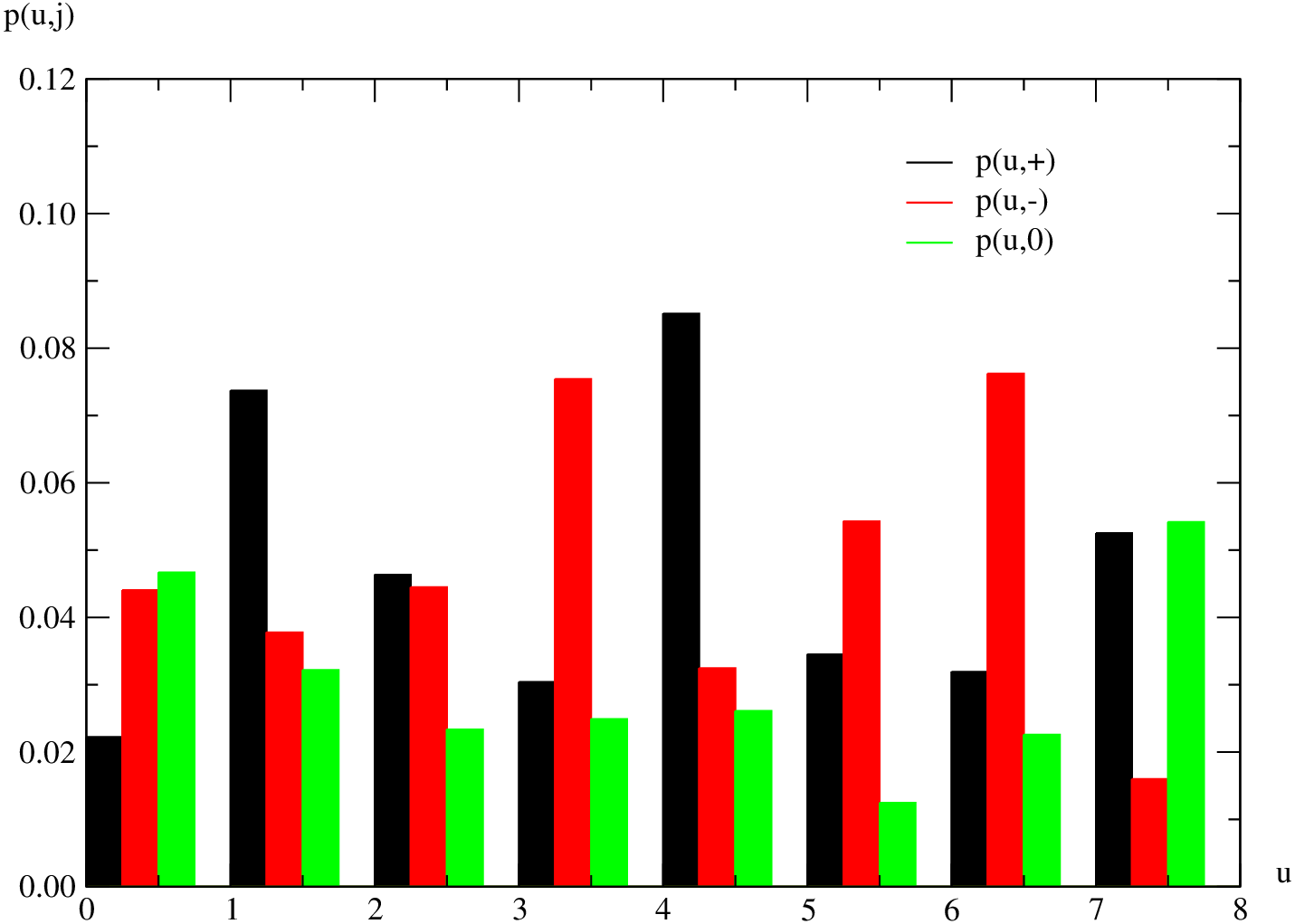}
\includegraphics[width=4.5cm]{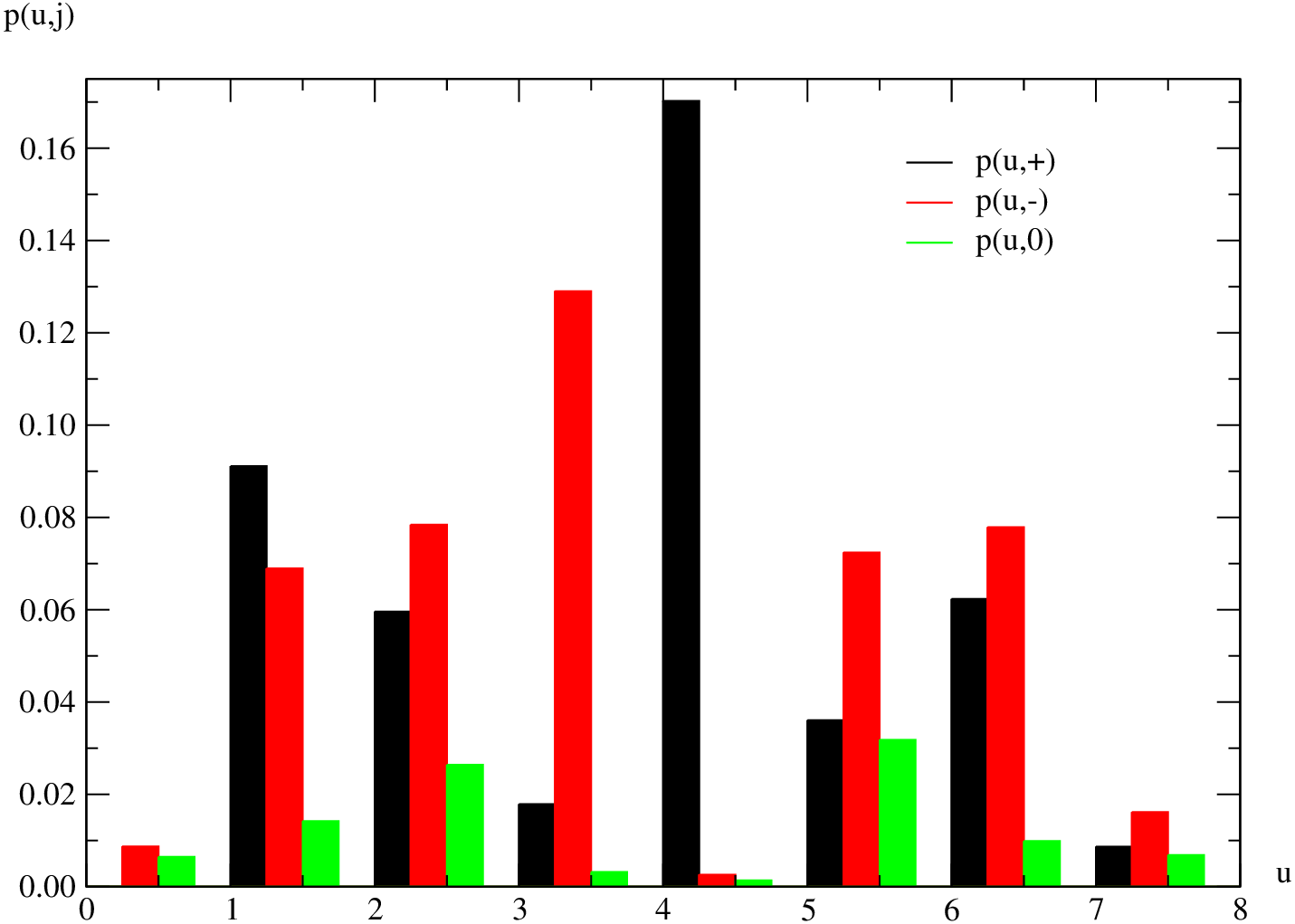}
\includegraphics[width=4.5cm]{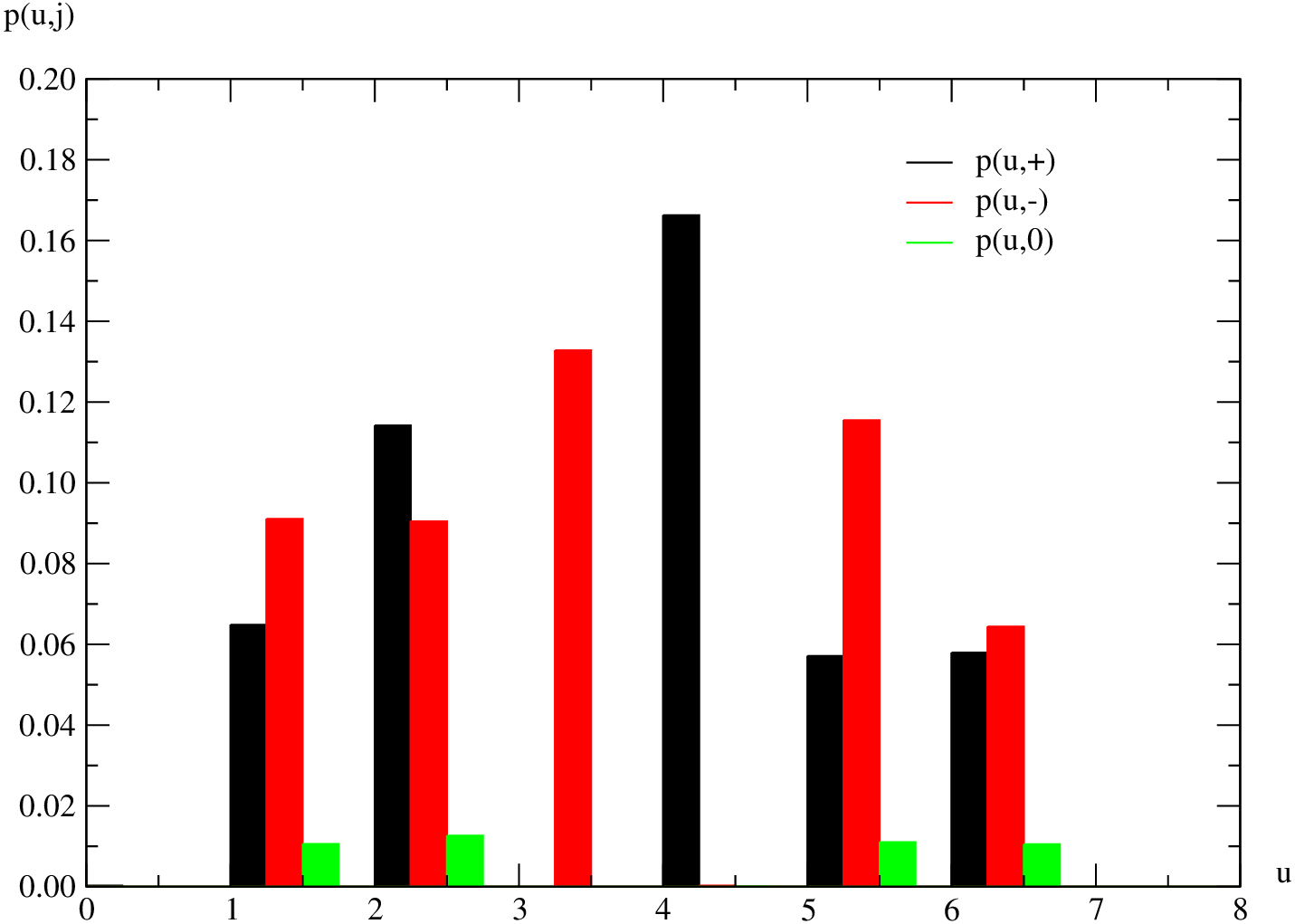}
\caption{\label{f3} Conditional probability $p(u,j)$ of the model
with linear price impact function for: (a) $N=50$, (b) $N=100$, (c)
$N=1000$ at $S=2$ and $m=3$.}
\end{figure}

We first compute the conditional probability $p(u,j)$, which is the
conditional probability to have positive, negative and zero price
change, denoted by $j=\pm1,0$, immediately following a specific
history $u$. In Fig.~\ref {f2}(a), (b) and (c), $p(u,j)$ for the
trading model with linear price impact function for $N=50,100,1000$
are plotted. In general, the histograms are not as flat as that of
the MG model \cite{sav99} which means the model has a bias of price
change conditional to a specific history. We observe that the
histogram for larger $N$ is less flat than the histogram for smaller
$N$, which means the price has a relative strong biased tendency at
large values of the parameter $N$. In Fig.~\ref {f3}(a), (b) and
(c), $p(u,j)$ for the trading model with square root price impact
function for $N=50,100,1000$ are plotted. The curves for the model
with square root price impact function is less flat than those of
the model with linear price impact function.

\begin{figure}[htb]
\centering
\includegraphics[width=6.5cm]{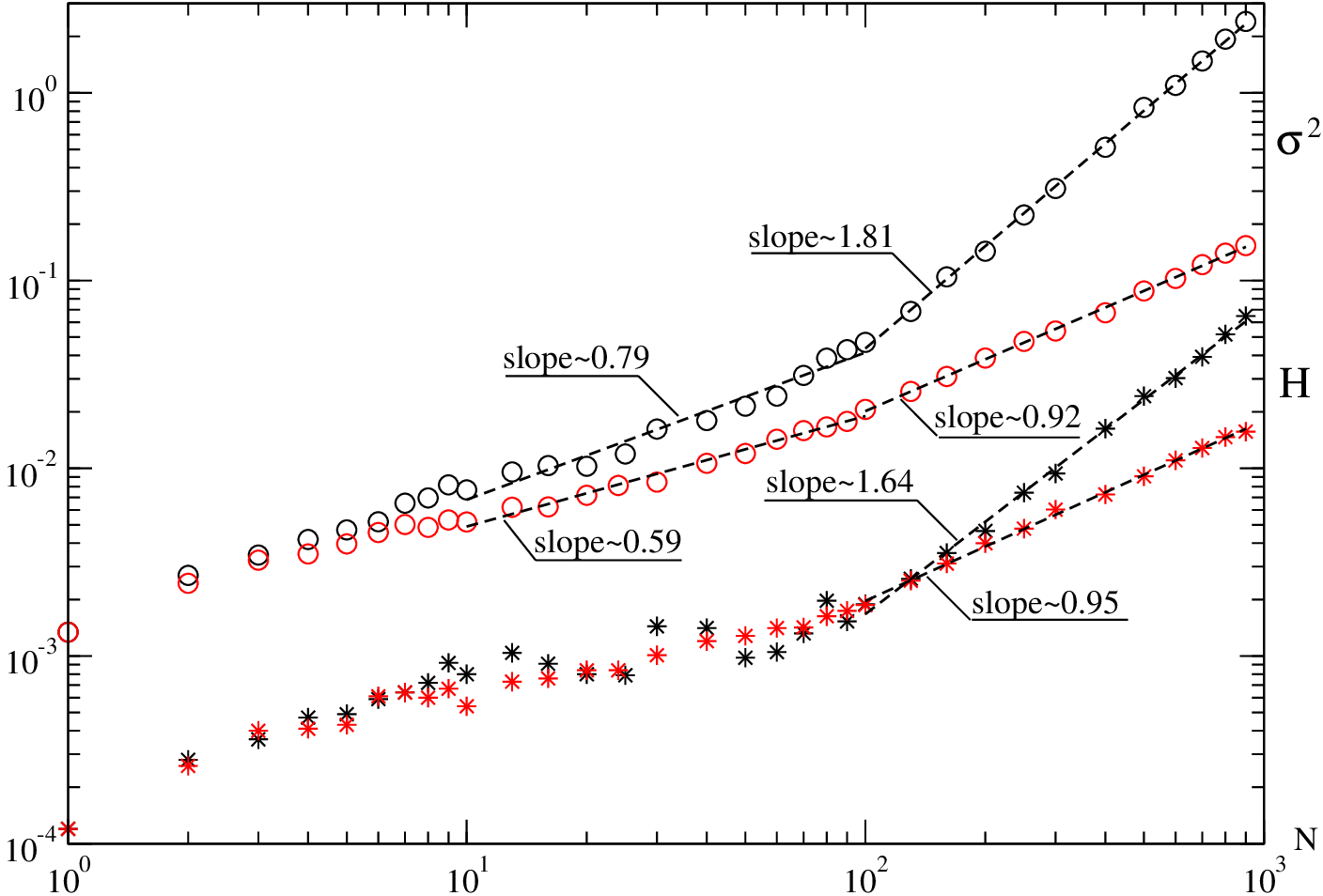}
\caption{\label{f4} Variance of returns $\sigma^{2}$ (circles) and
price impact $H$ (stars) for the trading models with linear and
square root price impact functions at $S=2$ and $m=3$, represented
by black and red symbols respectively. The results take average over
$100$ runs.}
\end{figure}

The variance of returns is a convenient reciprocal measure of the
market fluctuation, and it is defined as
\begin{equation}
\sigma^{2}=\frac{<r^2>-<r>^2}{P}. \label{e100}
\end{equation}
The smaller $\sigma^{2}$ is, the less the return fluctuates. In
Fig.~\ref {f4}, the variance $\sigma^{2}$ for the trading models
with linear and square root price impact functions are plotted,
denoted by black and red circles respectively. $\sigma^{2}$ for the
model with linear price impact function increases as the increase of
the parameter $N$, and obeys power laws with exponents $0.79$ for
$N\in[10,100)$ and $1.81$ for $N\in[100,1000)$. $\sigma^{2}$ for the
model with square root price impact function obeys power laws with
exponents $0.59$ for $N\in[10,100)$, and $0.92$ for
$N\in[100,1000)$. Compared with the model with linear price impact
function, the model with square root price impact function has a
smaller magnitude of the price fluctuation though it has a stronger
biased tendency of the price change.

To further understand the price return bias conditional to the
market states, we compute the average return conditional to a given
history defined as
\begin{equation}
H=\frac{\sum_{u} \langle r|u\rangle^2}{P}. \label{e110}
\end{equation}
In Fig.~\ref {f4}, $H$ for the trading models with linear and square
root price impact functions are also plotted, denoted by black and
red stars respectively. $H$ increases as the increase of the
parameter $N$, displaying a behavior similar to that of
$\sigma^{2}$. $H$ for the model with linear price impact function
seems saturated and fluctuates slightly for $N\in[10,100)$, then
obeys a nice power law with an exponent $1.64$ for $N\in[100,1000)$.
$H$ for the model with square root price impact function obeys a
power law with an exponent $0.95$ for $N\in[100,1000)$. These two
exponents are close to the half of the exponents of their price
impact functions, which indicates $H$ can be considered as an
approximate measure of the average price impact for large values of
the parameter $N$.

\begin{figure}[htb]
\centering
\includegraphics[width=6.5cm]{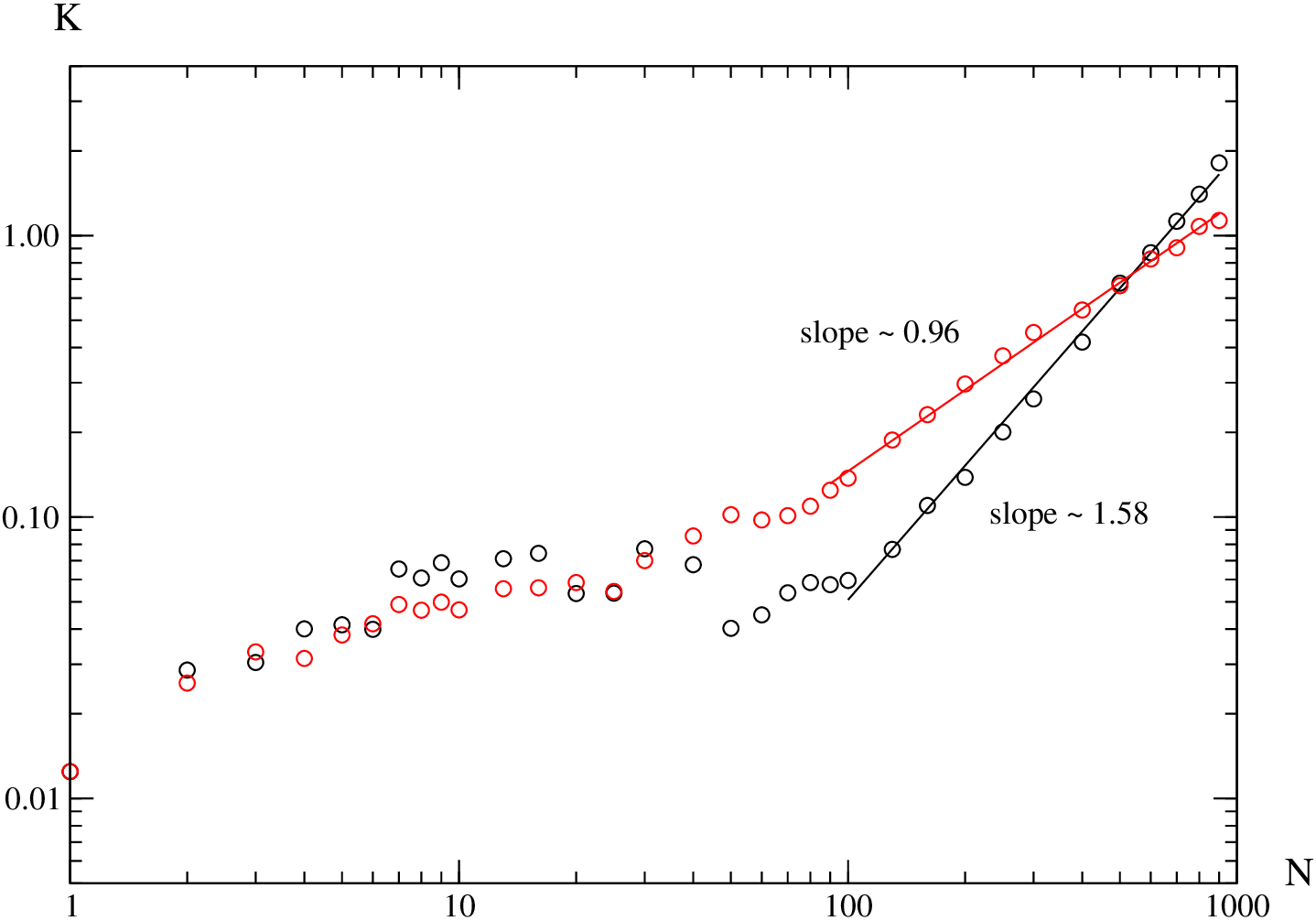}
\caption{\label{f5} Predictability $K$ for the trading models with
linear and square root price impact functions at $S=2$ and $m=3$,
represented by black and red circles respectively. The results take
average over $100$ runs.}
\end{figure}

Another important variable is the predictability that the traders
hope to exploit from the pair patterns
\begin{equation}
K=\frac{1}{P(P-1)}\sum_{\mu,\nu,\mu\neq\nu} \langle
r(t)|\mu\rightarrow\nu\rangle^2, \label{e120}
\end{equation}
where $\langle r(t)|\mu\rightarrow\nu\rangle$ stands for the average
price return per time step between the occurrence of $\mu$ at time
$t$ and the next occurrence of $\nu$. In Fig.~\ref {f5}, the
predictability $K$ for the model with linear and square root price
impact functions are plotted. For the model with linear price impact
function, $K$ increases at the early stage of the parameter $N$, and
shows an unclear behavior due to the unneglectable oscilation for
$N\in[10,100)$, then follows a power-law behavior with an exponent
$1.58$ for $N\in[100,1000)$. For the model with square root price
impact function, $K$ is a monotonously increasing function of the
parameter $N$, and obeys a power law with an exponent $0.96$ for
$N\in[100,1000)$. The model with square root price impact function
is more predictable than the model with linear price impact function
for $N\in[40,400]$, and tends to be less predictable if we further
increase the parameter $N$.

\begin{figure}[htb]
\centering
\includegraphics[width=6.5cm]{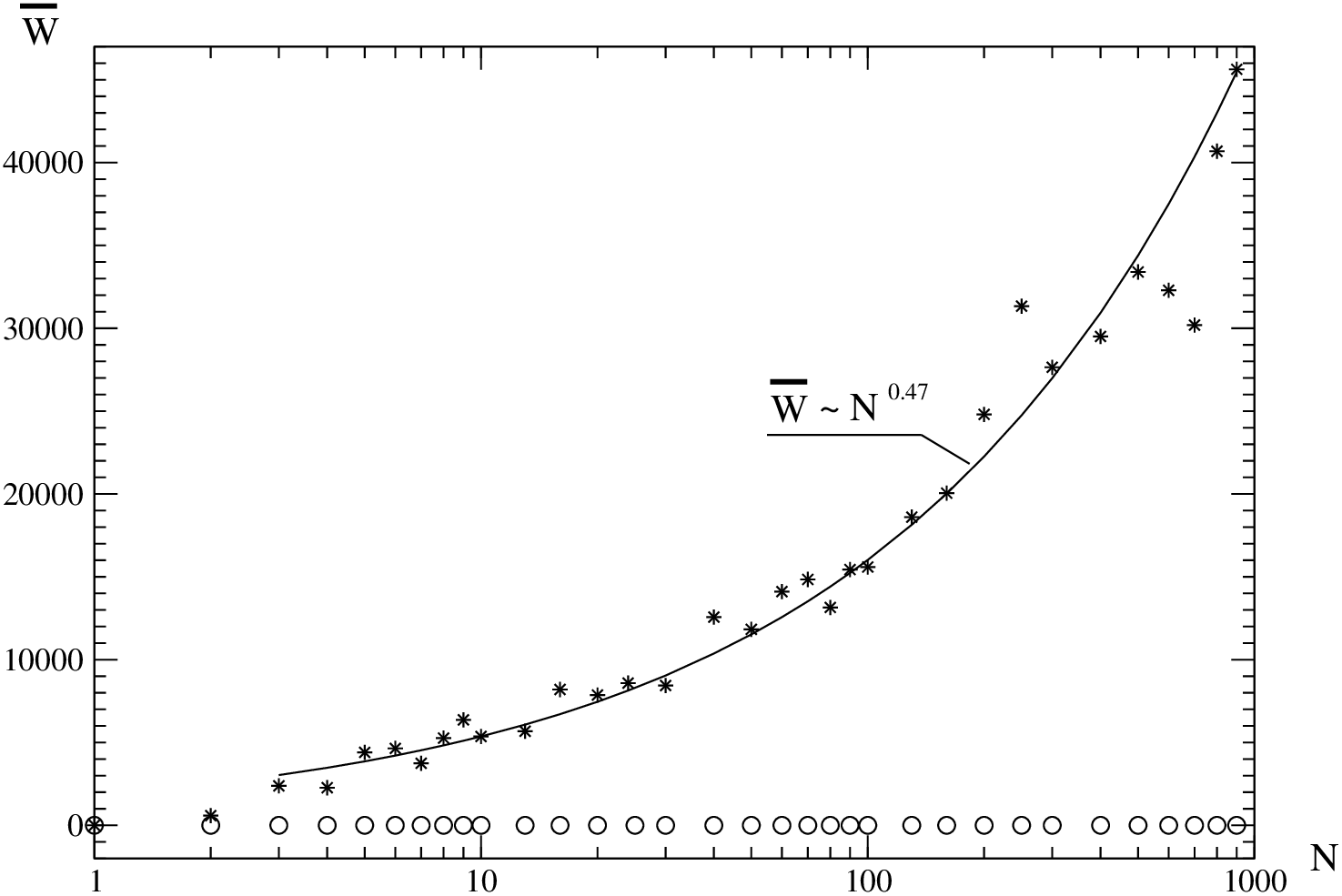}
\caption{\label{f6} Average wealth for each trader for the trading
models with linear and square root price impact functions at $S=2$
and $m=3$, represented by circles and stars respectively. The solid
line is the fitting curve $\overline{W}\sim N^{0.47}$.}
\end{figure}

We then consider the wealth of the traders. The average wealth for
each trader $\overline{W}=\frac{\sum_{i} W_i}{N}$ is calculated. In
Fig.~\ref {f6}, the average wealth for each trader for different
values of the parameter $N$ at $t'=10^6$ are plotted. The circles
and stars are for the trading models with linear and square root
price impact functions respectively. The average wealth for the
model with linear price impact function is close to zero independent
of the parameter $N$, which indicates that the system is a zero-sum
game. Remarkably, the model with square root price impact displys a
positive wealth sum: the curve increases as the increase of the
parameter $N$ and can be nicely fitted by a power law with an
exponent $0.47$.

\begin{figure}[htb]
\centering
\includegraphics[width=4.5cm]{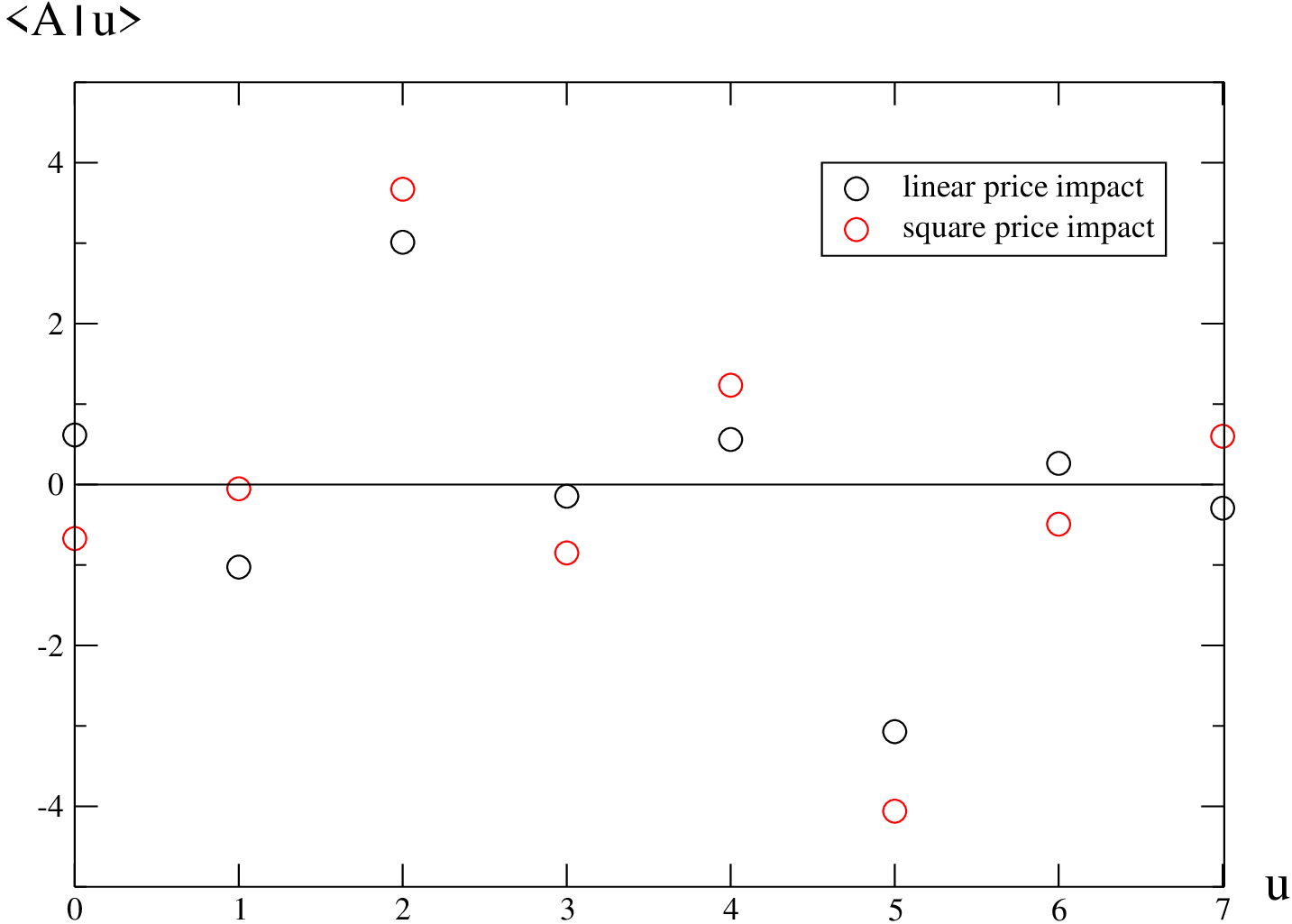}
\includegraphics[width=4.5cm]{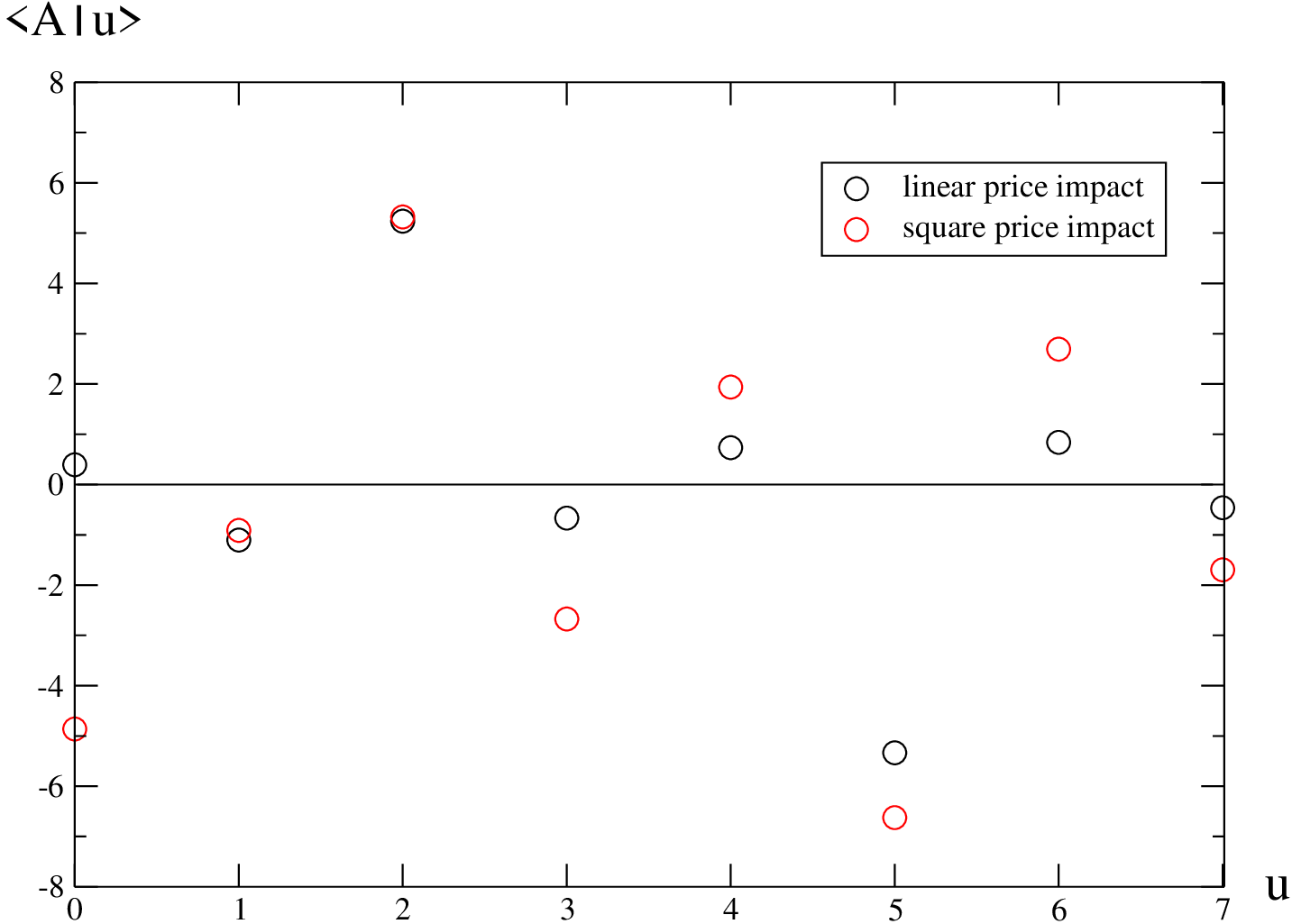}
\includegraphics[width=4.5cm]{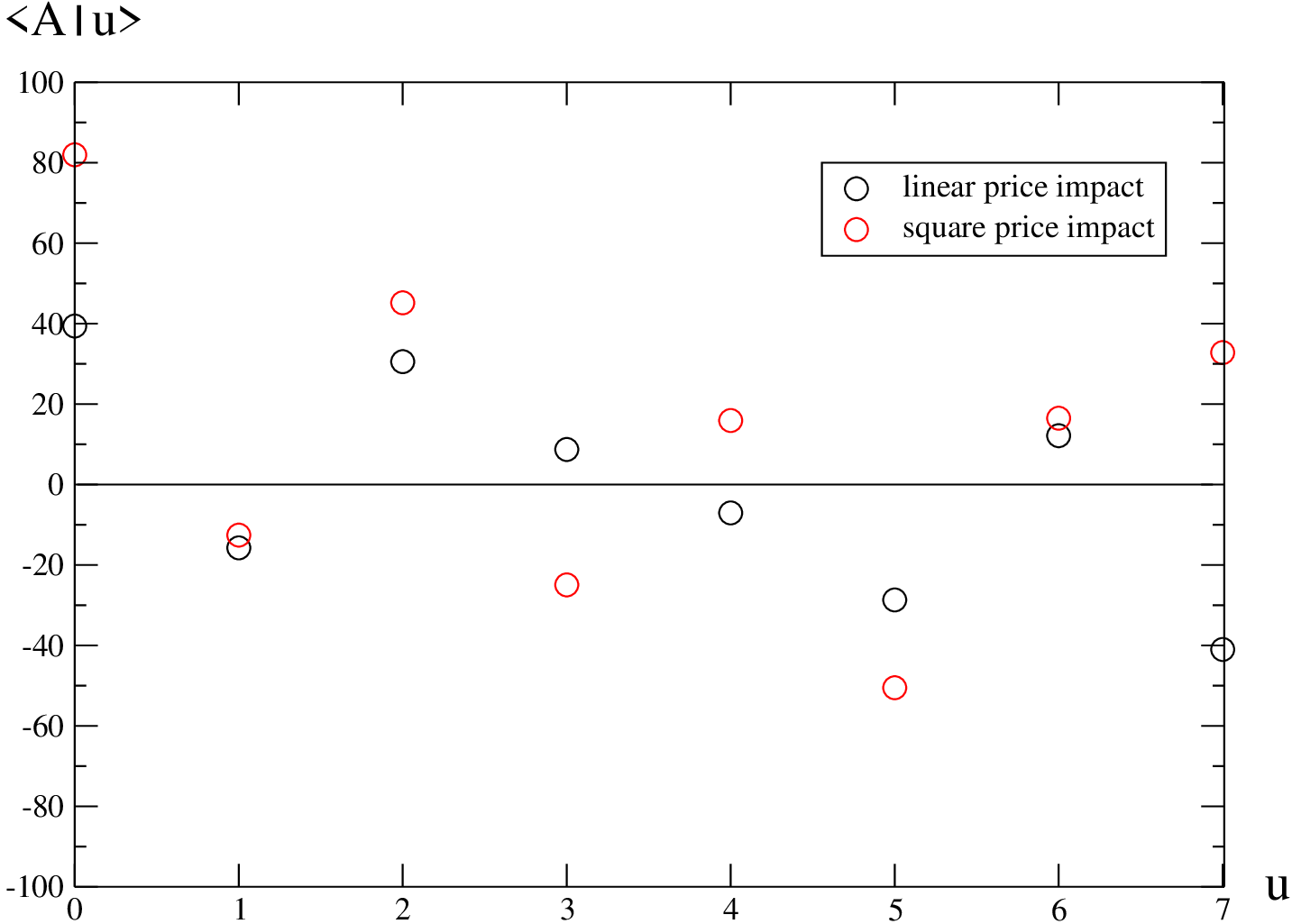}
\caption{\label{f7} Excess demand bias $\langle A|u \rangle$ for the
trading models with linear and square root price impact functions
for: (a) $N=50$, (b) $N=100$, (c) $N=1000$ at $S=2$ and $m=3$.}
\end{figure}

To understand why the trading model with linear price impact
function displays a zero sum and the trading model with square root
price impact function displays a positive sum, we further
investigate the average excess demand bias $\langle A|u \rangle$
conditional to a specific history $u$. In Fig.~\ref {f7} (a), (b)
and (c), $\langle A|u \rangle$ for the trading models with linear
and square root price impact functions (represented by black and red
circles respectively) for $N=50,100,1000$ are plotted. For the model
with linear price impact function, the sum of the trader's wealth
mainly depends on the cumulative access demand contributed by the
traders during the holding periods according to Eq. ({\ref {e80}}).
Assuming that the game visits each possible history with a equal
probability, so the sum over $\tau$ between two consecutive actions
of opening and closing a position could be substituted by the sum
over the possible history states between them. $\langle A|u \rangle$
for trading model with linear price impact function is symmetrically
distributed above and below zero for different states of history
$u$. $|\sum_u \langle A|u \rangle|$ displays a value close to zero.
Therefore, we observe a zero wealth sum.

The sum of the traders' wealth for the model with square root price
impact function mainly depends on the cumulative square root impact
of the access demand contributed by the traders during the holding
periods, supposing that the sum of the square root impact of the
excess demand at which the traders open and close their positions
equals zero, i.e., $\sum_{i}\sum_{t'_{i,1},t'_{i,2}}^{1\leq
t'_{i,1},t'_{i,2}\leq t'} [\frac{1}{2} sign(t'_{i,1})
\sqrt{A(t'_{i,1})}+ \frac{1}{2} sign(A(t'_{i,2}))
\sqrt{A(t'_{i,2})}]=0$, where $t'_{i,1}$ is the time trader $i$
opens a position and $t'_{i,2}$ is the time trader $i$ closes it.
$\langle A|u \rangle$ for the model with square root price impact
function is unsymmetrically distributed above and below zero.
$|\sum_u \langle A|u \rangle|$ displays a nonzero value obviously
larger than that of the model with linear price impact function,
e.g., the ratio of $|\sum_u \langle A|u \rangle|$ between two models
with square root and linear price impact functions is
$7.56,18.58,62.66$ for $N=50,100,1000$ . The larger the parameter
$N$ is, the more unsymmetrical the distribution of $\langle A|u
\rangle$ is. Some traders have certain strategies which can help
them to effectively exploit the information of the biased excess
demand from the patterns, and they make profits from these
high-performed strategies. They have wealth greater than zero while
other traders have an average wealth close to zero. This may leads
to a positive sum for the model with square root impact function.

\begin{figure}[htb]
\centering
\includegraphics[width=6.5cm]{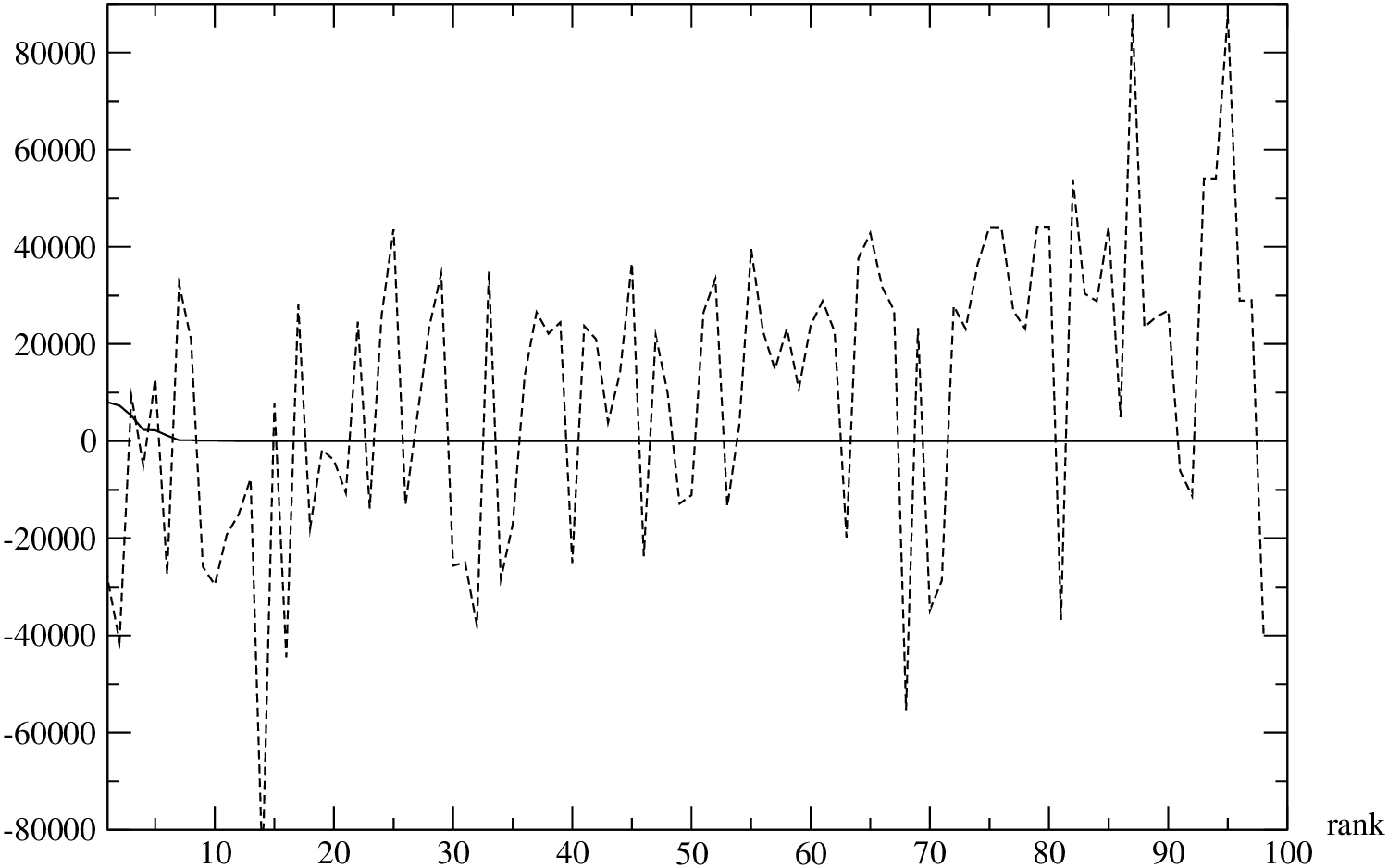}
\includegraphics[width=6.5cm]{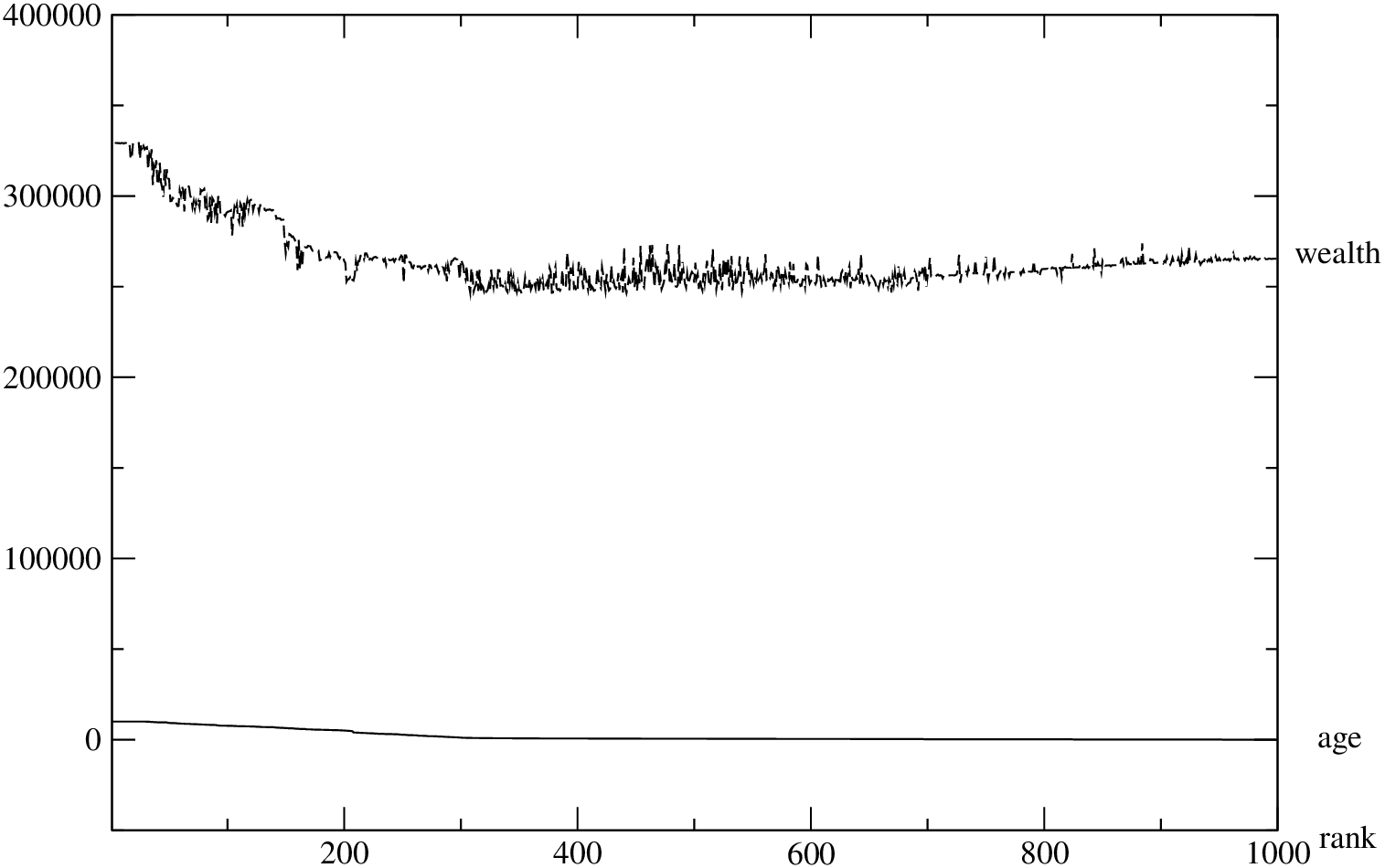}
\caption{\label{f8} (a) Wealth distribution of traders and change
frequency of adopted strategies for the model with square root price
impact function for $N=100$ at $S=2$ and $m=3$, represented by
dashed and solid lines separately. (b) Wealth distribution and age
of traders for the trading model with evolution for $N=1000$ at
$S=2$ and $m=3$, represented by dashed and solid lines separately.}
\end{figure}

In Fig.~\ref {f8} (a), the wealth distribution of the traders for
the model with square root price impact function for $N=100$ is
plotted according to the rank of their change frequency of the
adopted strategies. We observe that the traders who change their
strategies more frequently have less wealth. Some traders keep using
their high-performed strategies to make more profits, while the
others who do not have these strategies always shift among their
strategies and have less wealth. Especially for some singular runs
with attractors, we observe that some traders keep using certain
high-performed strategies and the others shift among their
strategies at the early stage of the evolution and eventually
withdraw from the market and thus the system is stuck in a string of
period history states. For the model with linear price impact
function, the wealth distribution shows a similar behavior but
displays a zero sum.

\section{Trading Model with Evolution}

MG-based models with dynamic evolution have been studied in Refs.
\cite{sys03,li00,cha01a}. For example, in Refs. \cite{sys03,li00}
the traders can change their strategies with poor performances. In
this trading model, we assume that the worst trader can be driven
out of the market following Ref \cite{cha01a}. Every 100 time steps
the trader who has the lowest wealth is washed out and a new trader
with new randomly selected strategies is generated. The wealth of
the new trader is set to be the average wealth of the traders at
that moment. We use the square root price impact function of the
real markets in this evolutionary trading model. With this evolution
mechanism, the price evolution becomes more continuously, and seems
to be similar to that of the real markets as it is shown in
Fig.~\ref {f1}.

\begin{figure}[htb]
\centering
\includegraphics[width=6.5cm]{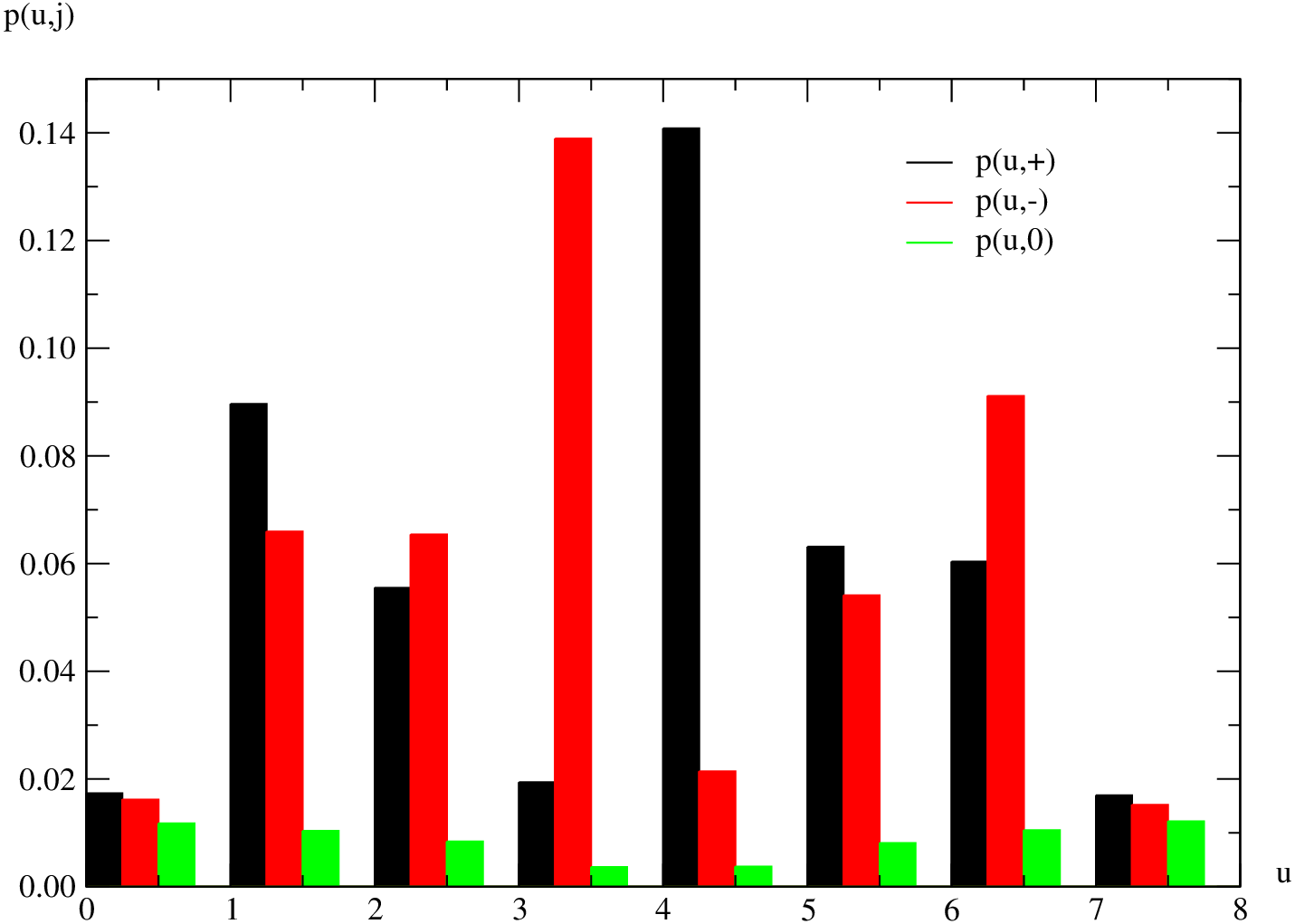}
\includegraphics[width=6.5cm]{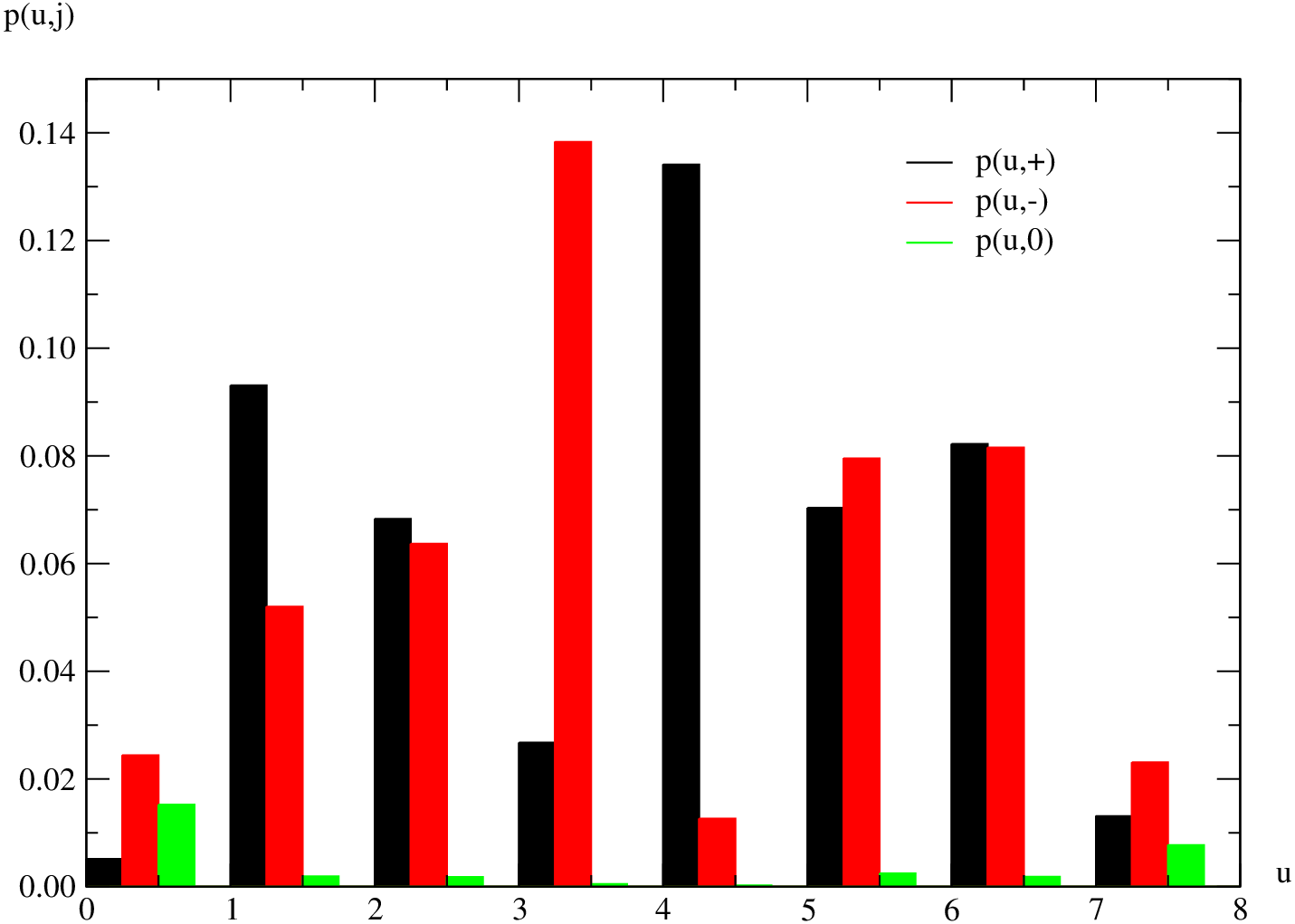}
\caption{\label{f9} Conditional probability $p(u,j)$ of the trading
model with evolution for: (a) $N=100$, (b)$N=1000$ at $S=2$ and
$m=3$.}
\end{figure}

The conditional probability $p(u,j)$ for different values of the
parameter $N=100,1000$ for this evolutionary trading model are
plotted in Fig.~\ref {f9} (a) and Fig.~\ref {f9} (b). It seems that
the histograms of the model with evolution are more flat than those
of the model without evolution for the same values of the parameter
$N$. The introduction of the elimination mechanism leads to a
relatively week biased tendency of the price change. $\sigma^{2}$,
$H$ and $K$ for the evolutionary model are also effectively
decreased. In general, the elimination mechanism breaks down the
domination of the strategies highly performed in the trading model
and makes the price fluctuations relatively symmetric.

\begin{figure}[htb]
\centering
\includegraphics[width=6.5cm]{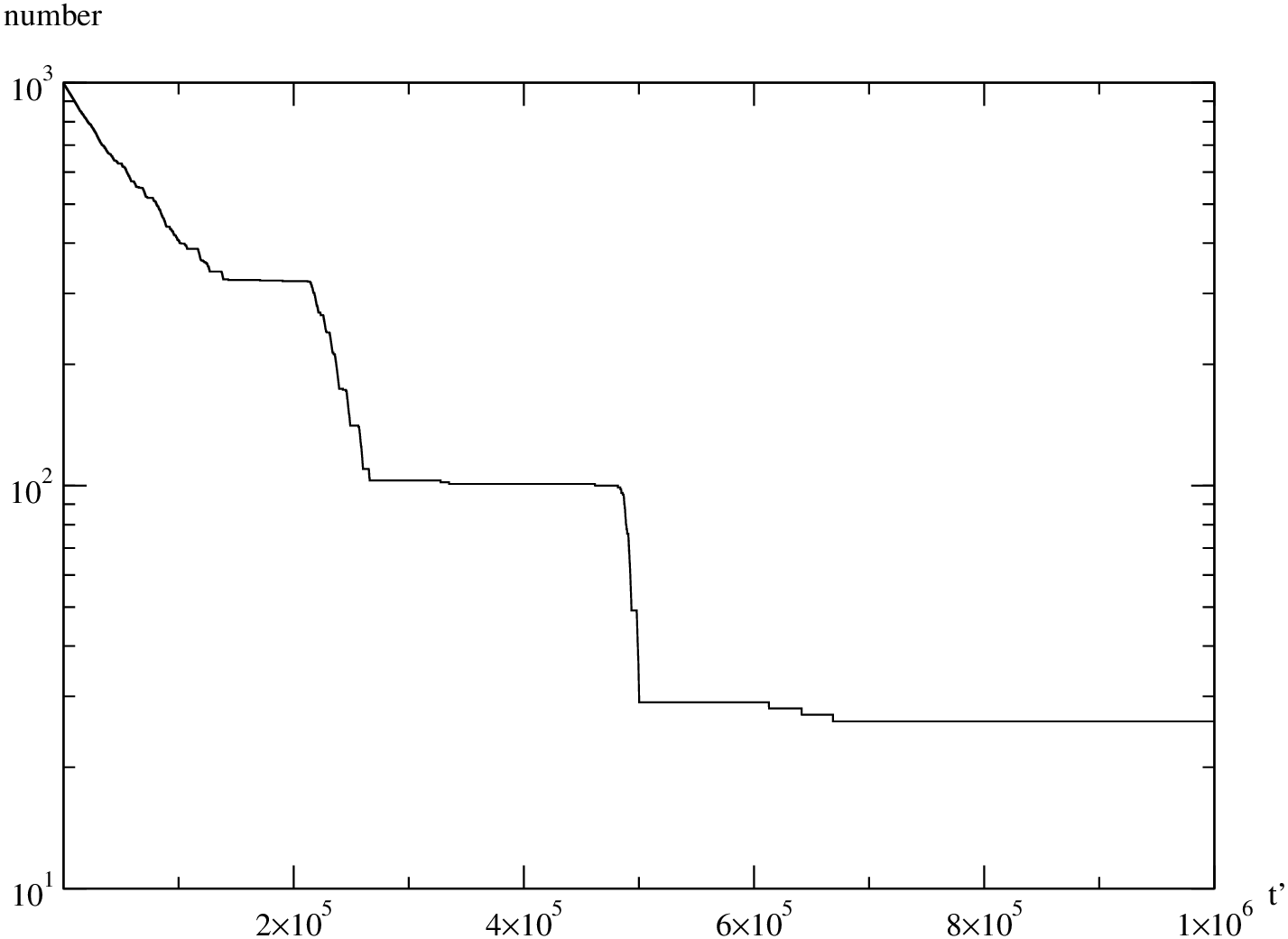}
\caption{\label{f10} Number of traders still survive evolves with
time $t'$ for the trading model with evolution at $S=2$, $m=3$, and
$N=1000$.}
\end{figure}

\begin{figure}[htb]
\centering
\includegraphics[width=6.5cm]{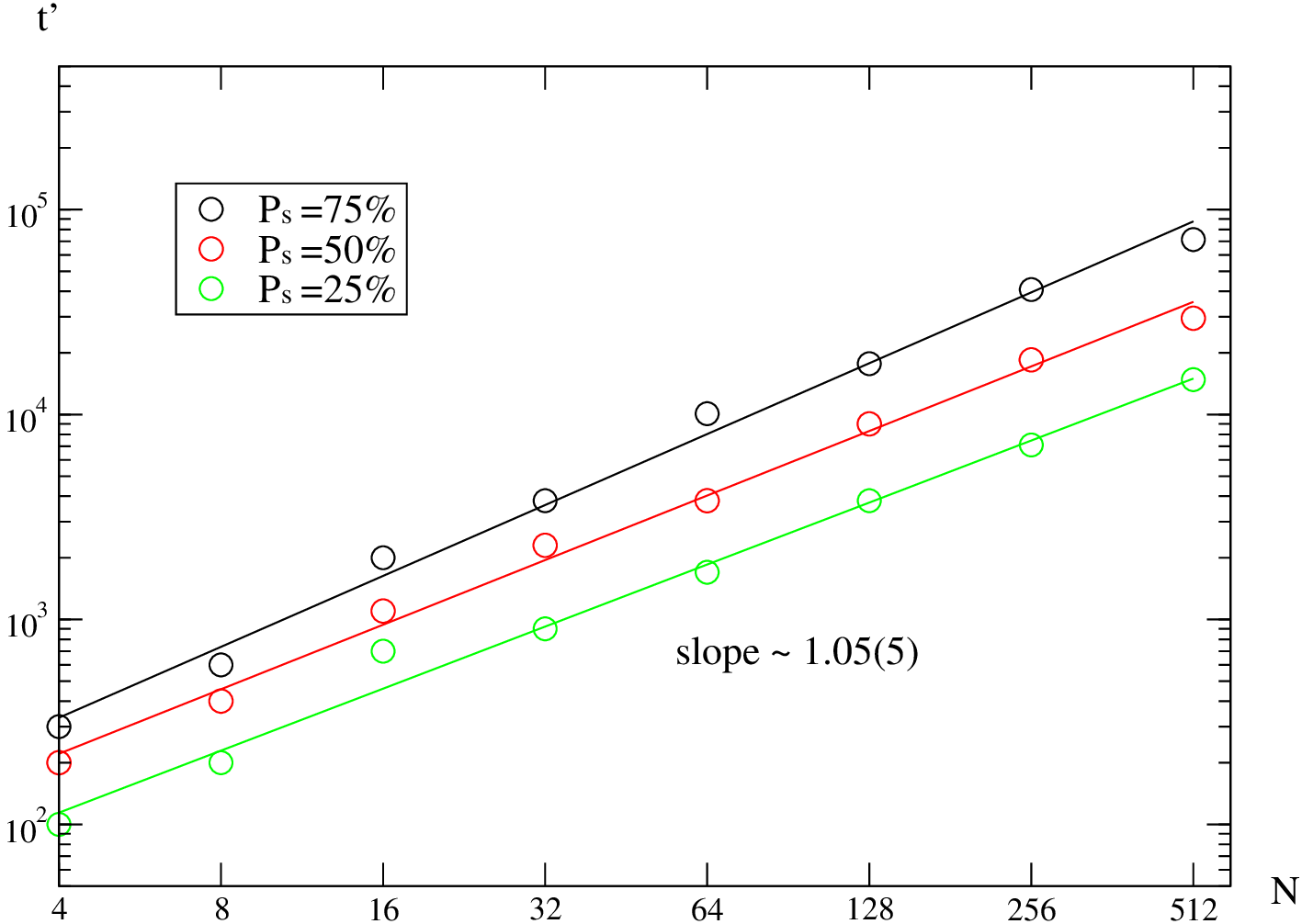}
\caption{\label{f11} Time $t'$ at which $P_s$ percentage of the
traders are washed out for the trading model with evolution at $S=2$
and $m=3$.}
\end{figure}

In Fig.~\ref {f10}, the number of the traders still survive evolved
with time $t'$ is plotted. For a large number of traders, e.g.,
$N=1000$, those traders who have the strategy (2,5) could survive
for quite a long time and are finally washed out one after another
in a short time region. There is no high-performed strategy always
keep winning after we introduce the elimination mechanism. We also
observe that the time at which the traders are washed out mainly
depends on the parameter $N$. Fig.~\ref {f11} shows the time $t'$ at
which $P_s$ percentage of the traders are washed out as a function
of the parameter $N$. $t'$ increases as the increase of the
parameter $N$, and obeys a power law with an exponent close to
$1.05$ not much different for different values of the parameter
$P_s=25\%,50\%,75\%$.

\begin{figure}[htb]
\centering
\includegraphics[width=6.5cm]{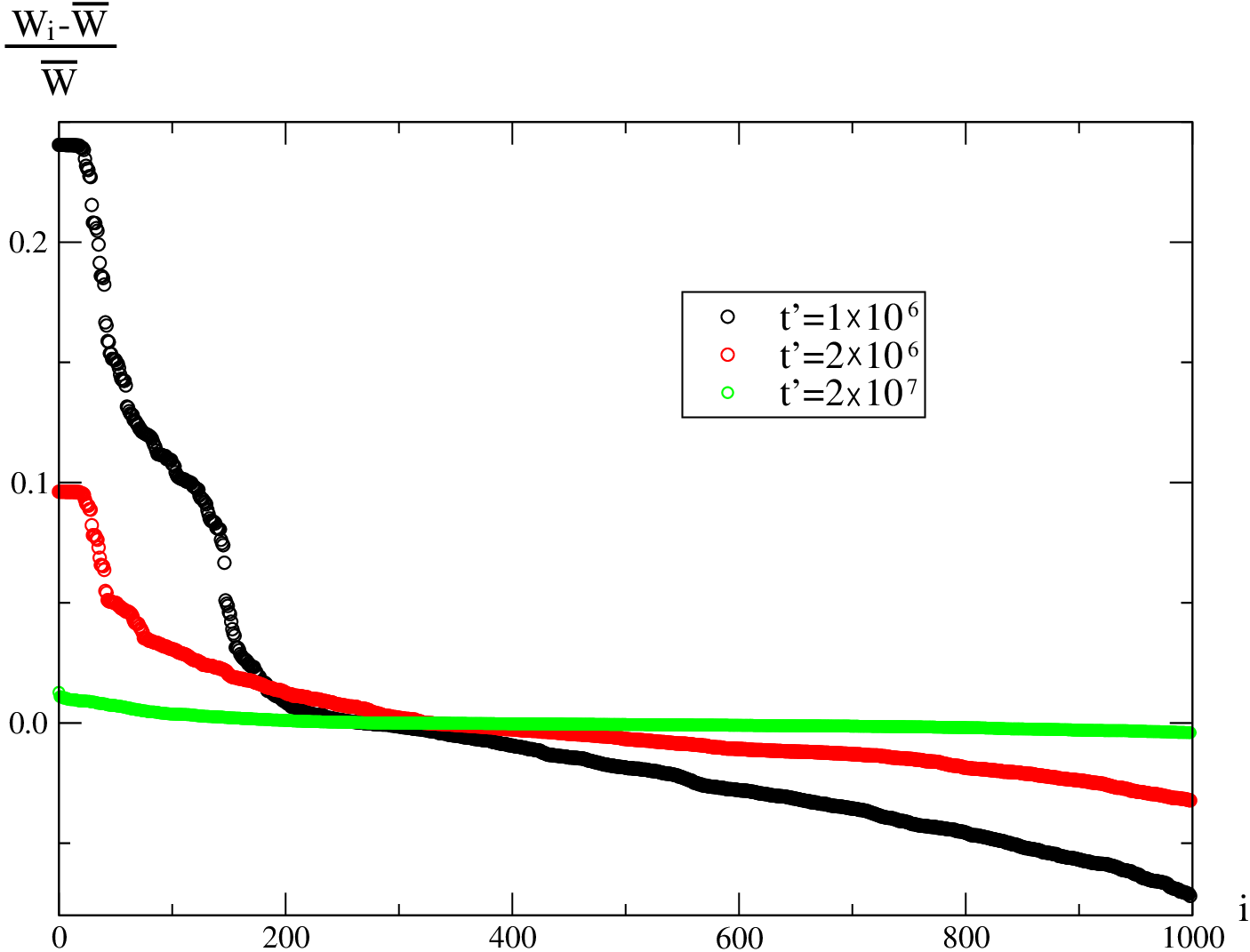}
\caption{\label{f12} Relative wealth distribution
$(W_i{t'}-\overline{W})/\overline{W}$ at different time
$t'=1\times10^6, 2\times10^6, 2\times10^7, 5\times10^7$ for the
trading model with evolution at $S=2$, $m=3$ and $N=1000$.}
\end{figure}

Let's take a look at how the traders' wealth evolve before they are
washed out. In Fig.~\ref {f12}, the distribution of the relative
wealth $(W_i(t')-\overline{W})/\overline{W}$ at different time
$t'=1\times10^6, 2\times10^6, 2\times10^7$ are plotted, where $i$ is
the rank of the trader's wealth and $\overline{W}$ is the average
wealth of the traders at time $t'$. For $t'=1\times10^6$, the
relative wealth is not continuous distributed among the traders.
Those traders who have higher wealth are clustered in different
groups. For $t'=2\times10^6$, the distribution remains similar, but
the relative wealth difference between the rich traders and the poor
traders is not so large and thus the curve becomes more flat. At the
time just before all the traders are washed out, e.g.,
$t'=2\times10^7$, the relative wealth distribution becomes even more
flat.

\begin{figure}[htb]
\centering
\includegraphics[width=6.5cm]{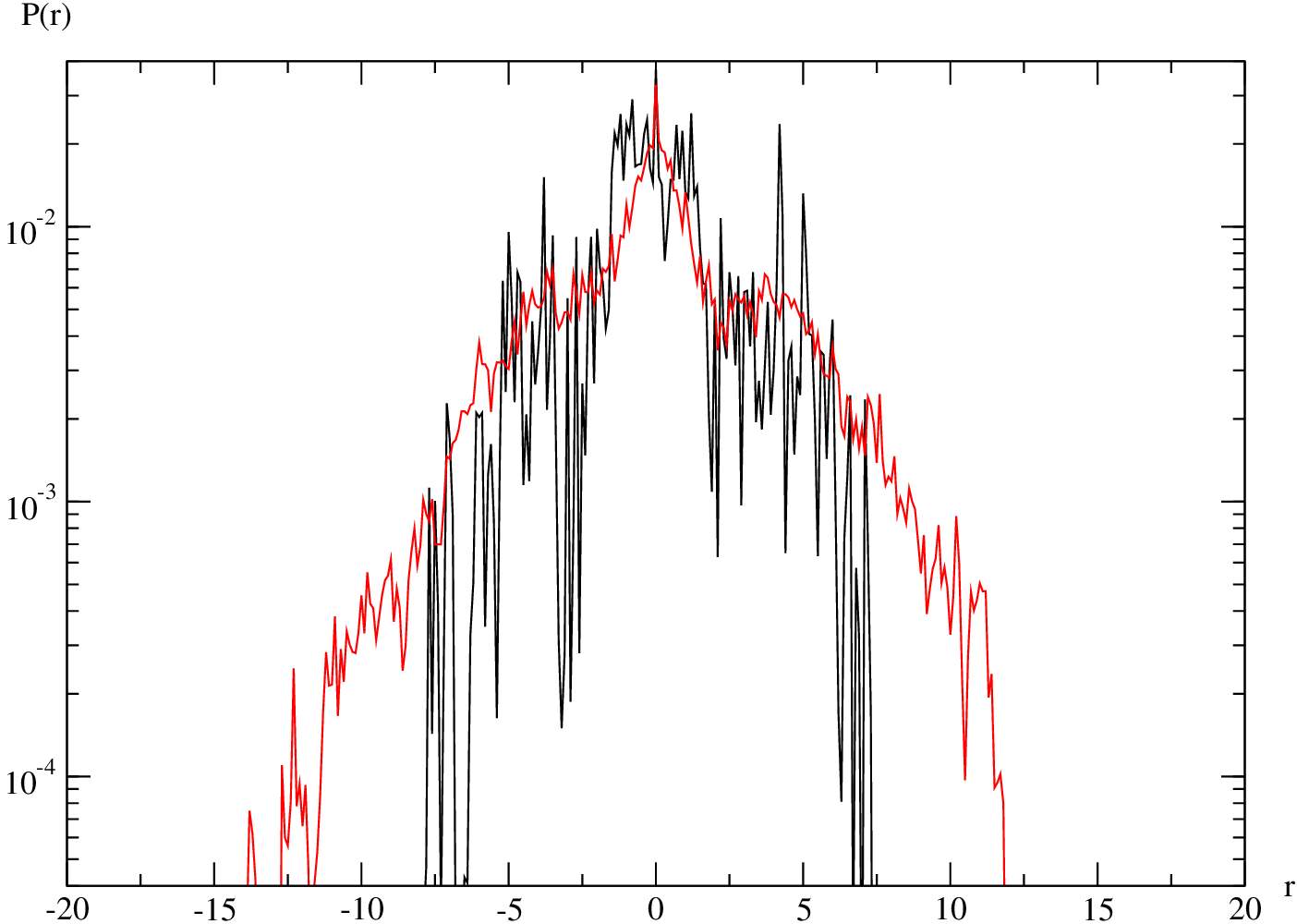}
\caption{\label{f13} Return distribution of the trading models with
and without evolution for $S=2$, $m=3$ and $N=1000$, represented by
red and black lines separately.}
\end{figure}

The wealth distribution of the traders ranked by their age (survival
time) for the trading model with evolution is plotted in Fig.~\ref
{f8} (b). We observe that all the traders have positive wealth due
to the evolution mechanism. Those elder traders have relatively more
wealth than those younger traders, and those traders newly generated
have an average wealth among them. In Fig.~\ref {f13}, the return
distributions of the models with and without evolution for a single
run are plotted. Though the return distributions of both models
decay exponentially, the model with evolution has a tail fatter than
the model without evolution.

\section{Trading Model with Other Types of Traders}

We introduce the traders who have the strategies the same as those
in the MG model \cite{cha97,zha98}, which give the predictions for
all the probable history status, to the trading model with square
root price impact function but without evolution. Each of these
newly introduced traders has the same number of $S$ randomly
selected strategies, and they trade at each time step using their
best strategies according to the pattern (or history) shared by all
the traders. Let $N_t$ and $N_m$ be the number of the traders who
have the pair pattern strategies and who have the strategies the
same as those in the MG model. Then the excess demand is defined as
$A(t)=\sum_{i=1}^{N_t+N_m} a_i(t)$, and the price dynamics remains
the same as Eq. (\ref {e90}). The score of the pair pattern strategy
takes the same update form as Eq. ({\ref {e50}}), and the score of
the MG's strategy is updated as
$U_{i,s}(t'+1)=U_{i,s}(t')-a_i(t')(P(t'+1)-P(t'))$, $i=1,...,N_m$.

\begin{figure}[htb]
\centering
\includegraphics[width=6.5cm]{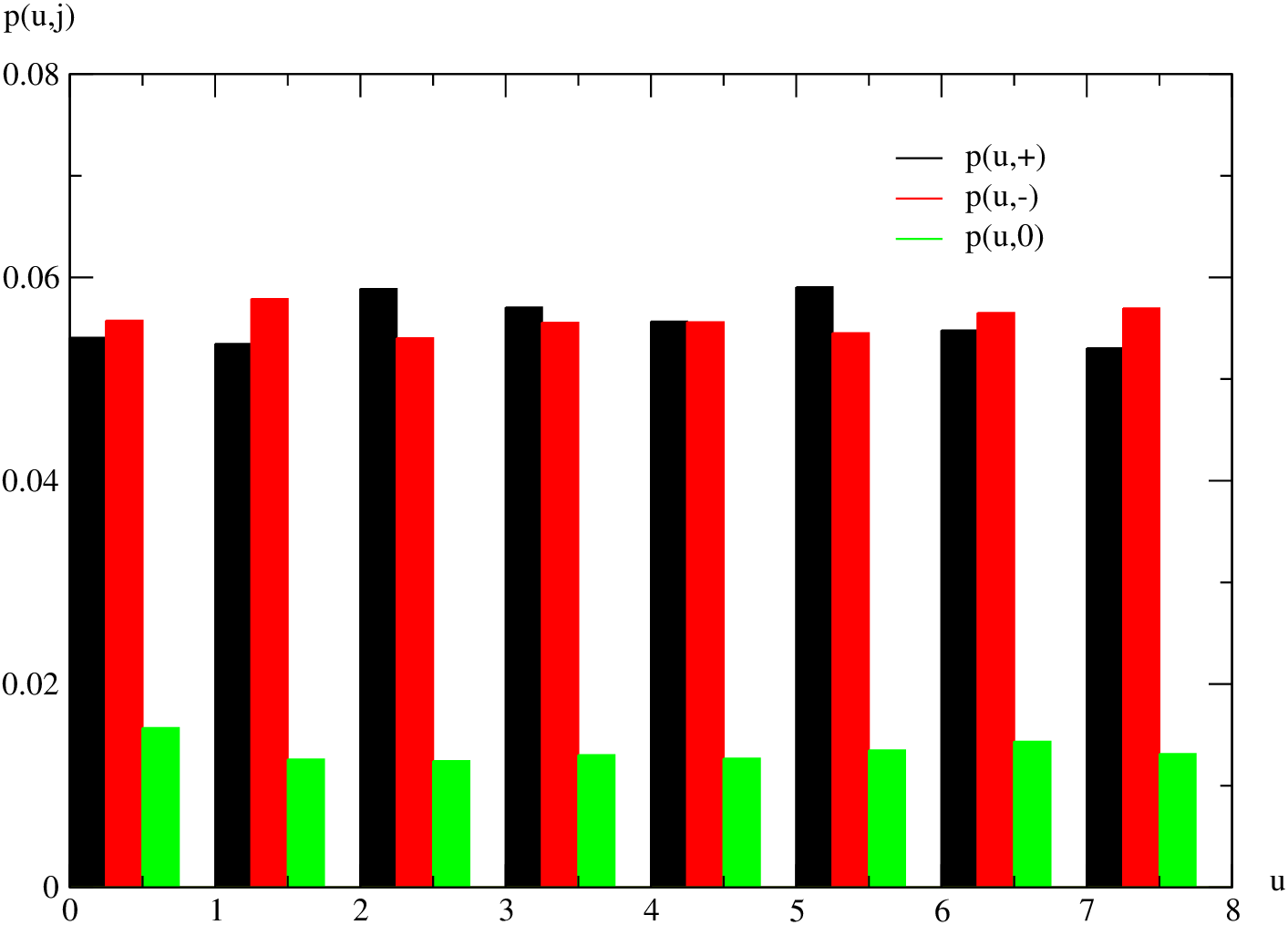}
\caption{\label{f14} Conditional probability $p(u,j)$ for the
trading model with mixed population $N_t=100$ and $N_m=25$ at $S=2$
and $m=3$.}
\end{figure}

\begin{figure}[htb]
\centering
\includegraphics[width=6.5cm]{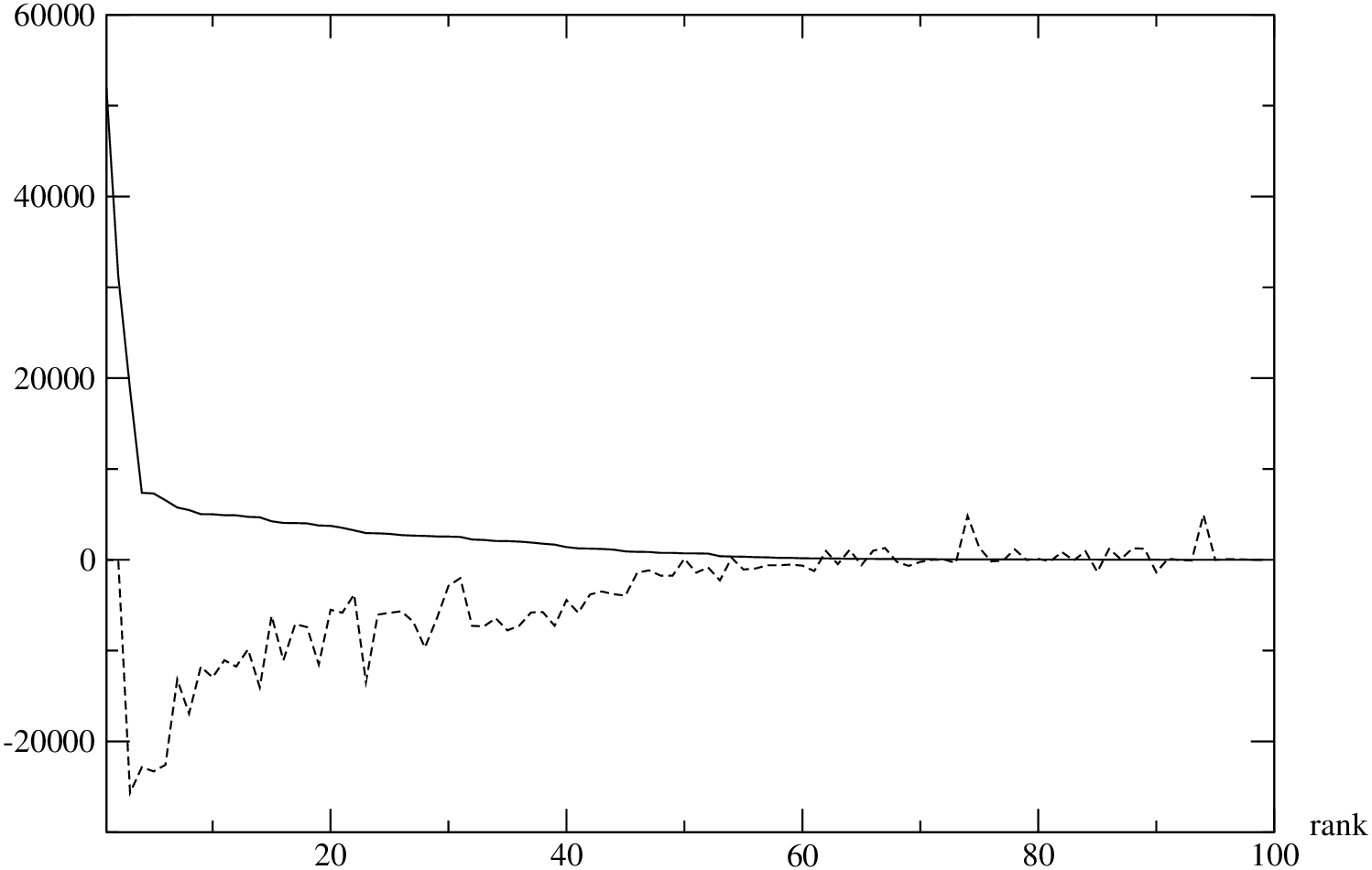}
\includegraphics[width=6.5cm]{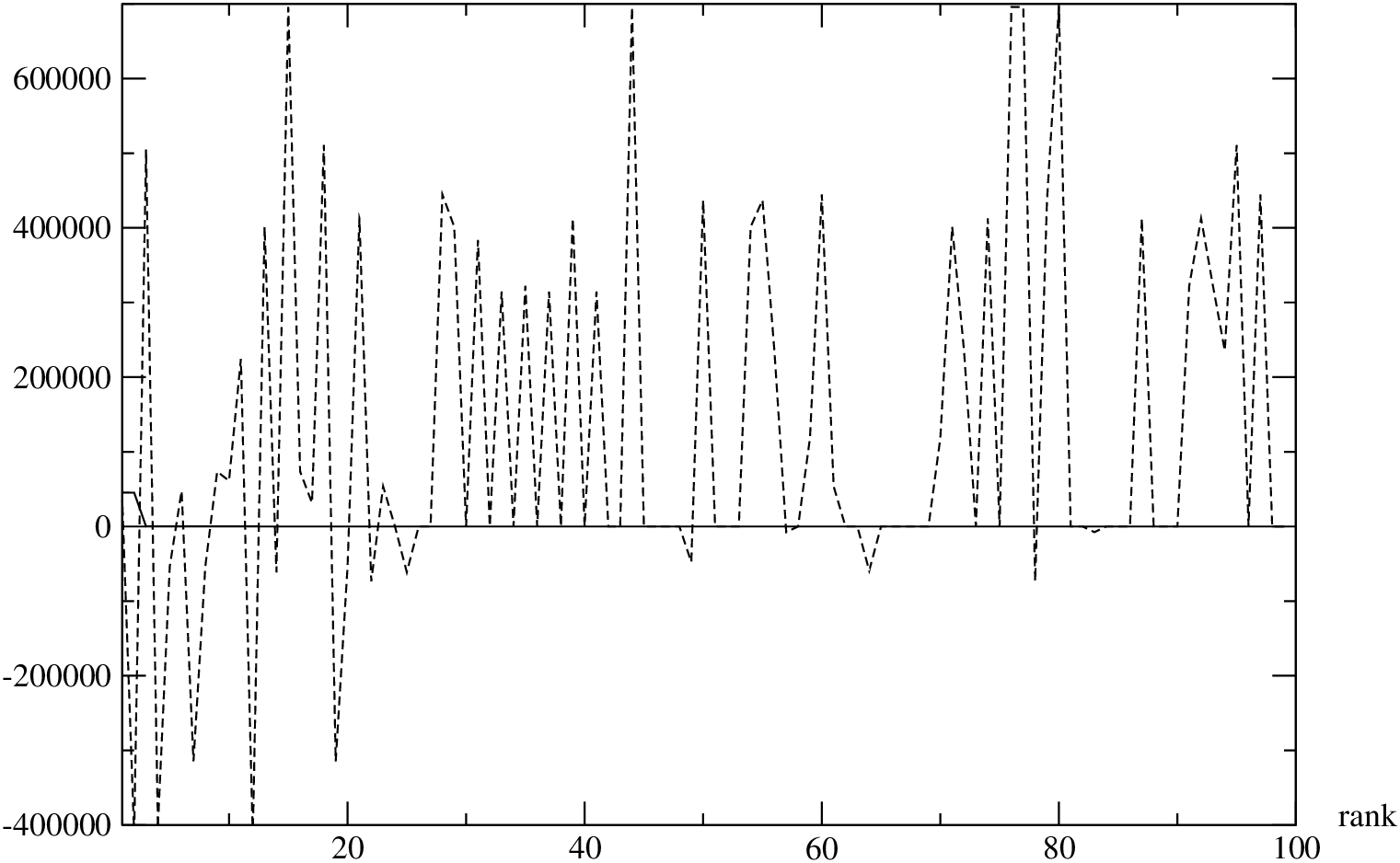}
\caption{\label{f15} Wealth distribution and the change frequency of
the adopted strategies of the traders who have the pair pattern
strategies for the trading model with mixed population: (a)
$N_t=100$ and $N_m=25$, (b) $N_t=100$ and $N_p=100$ (the number of
the traders known as producers) at $S=2$ and $m=3$, represented by
dashed and solid lines separately.}
\end{figure}

The conditional probability $p(u,j)$ and the wealth distribution of
the traders ranked by their change frequency of the adopted
strategies for $N_t=100$ and $N_m=25$ are plotted in Fig.~\ref {f14}
and Fig.~\ref {f15} (a). The number of the traders effectively trade
at each time step ${N_t}_{eff}:{N_m}_{eff}=1:1$. The histogram of
the conditional probability becomes much more flat than that of the
model has pure traders who have the pair pattern strategies, and the
sum of the wealth of the traders who have the pair pattern
strategies tends to be negative. The traders who have the MG's
strategies distinctly affect the behavior of the traders who have
the pair pattern strategies.

\begin{figure}[htb]
\centering
\includegraphics[width=6.5cm]{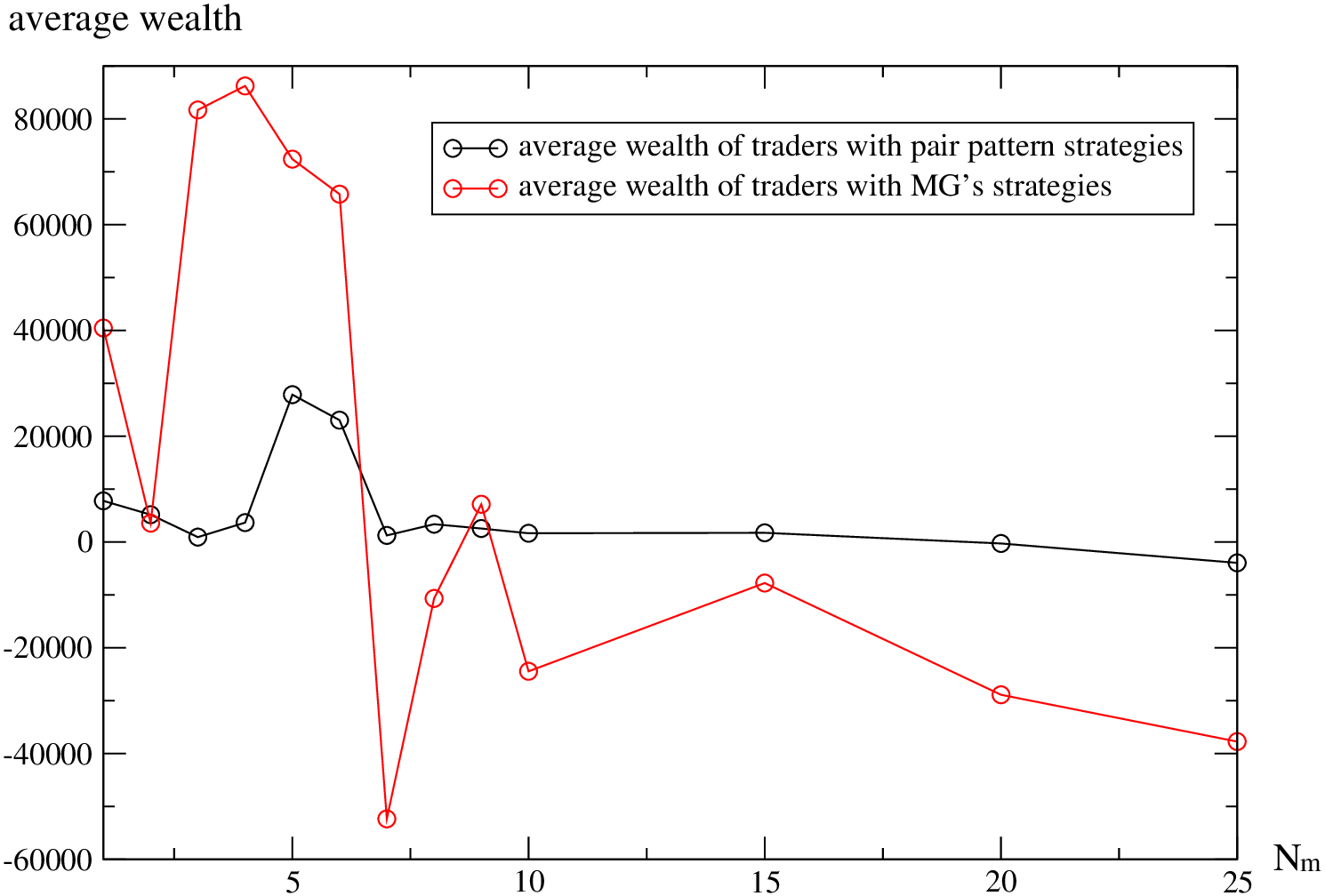}
\includegraphics[width=6.5cm]{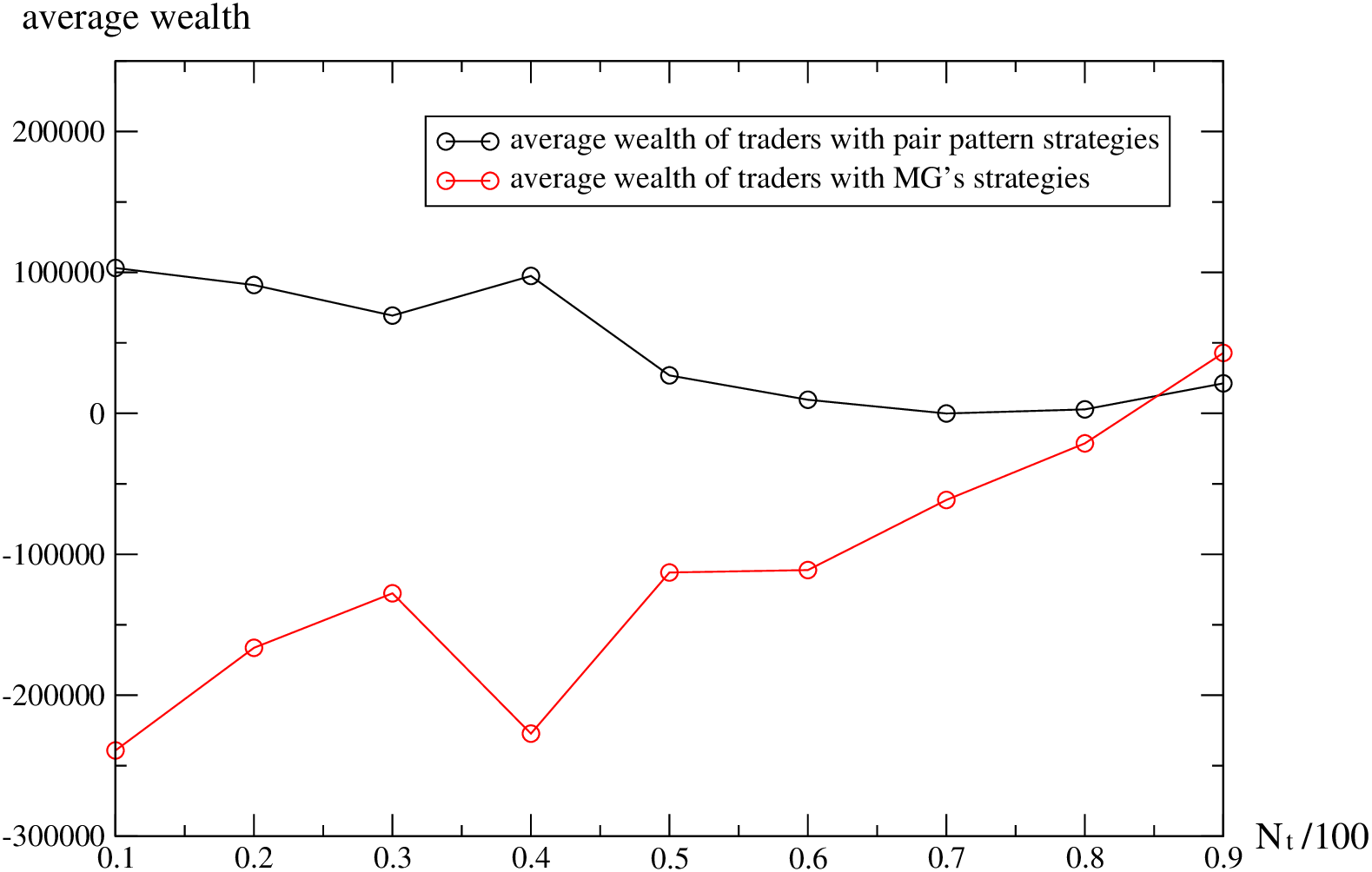}
\caption{\label{f16} Average wealth of the traders who have the pair
pattern strategies and average wealth of the traders who have the
MG's strategies for the trading model with mixed population: (a)
$N_m$ ranging from 1 to 25 at fixed $N_t=100$, (b) different
proportion of $N_t/100$ at fixed $N_t+N_m=100$ for $S=2$ and $m=3$.}
\end{figure}

We fix the number of the traders who have the pair pattern
strategies $N_t=100$, and increase the number of the traders who
have the MG's strategies one by one and see how the traders' wealth
behave. In Fig.~\ref {f16} (a), the average wealth of the traders
who have the pair pattern strategies and the traders who have the
MG's strategies for $N_m$ ranging from 1 to 25 at fixed $N_t=100$
are plotted. For a small $N_m$, the average wealth of the traders
who have the MG's strategies is positive and larger than that of the
traders who have pair pattern strategies. Few number of the traders
dominant the game, and are fed by the rest majority traders who have
positive wealth sum. Interestingly, at $N_m \sim 5$ the average
wealth of both types of traders reaches a maximum. Those two types
of traders seem to have a certain state of corporation. As we
further increase $N_m$, the average wealth of the traders who have
the MG's strategies tends to be negative and smaller than that of
the traders who have pair pattern strategies.

Another case that the total number of the traders is fixed is also
considered, i.e., $N_t+N_m=100$. We change the proportion of $N_t$
to the total number of the traders. In Fig.~\ref {f16} (b), the
average wealth of both types of traders are plotted for fixed
$N_t+N_m=100$. The average wealth of the traders who have the MG's
strategies increases with the increase of the proportion of
$N_t/100$, while the average wealth of the traders who have the pair
pattern strategies decreases with the increase of the proportion of
$N_t/100$. For a small $N_t/100$, the average wealth of the traders
who have the MG's strategies is negative and smaller than that of
the traders who have the pair pattern strategies, and tends to be
positive and larger than that of the traders who have the pair
pattern strategies for $N_t/100 \sim 0.9$. That is consistent with
the result we obtained for fixed $N_t=100$ shown in Fig.~\ref {f16}
(a).

We also introduce the traders who only have one MG's strategy known
as "producers" \cite{cha01} to the trading model with square root
price impact function but without evolution. As it is shown in
Fig.~\ref {f15} (b), the introduce of this type of traders can
effectively increase the wealth of the traders who have the pair
pattern strategies. Most of the traders who have the pair pattern
strategies have positive wealth, but the wealth distribution is more
fluctuated than the model with the traders who have the MG's
strategies shown in Fig.~\ref {f15} (a).

\section{Conclusion}

In summary, trading model with pair pattern strategies evolved with
middle price is introduced. Both linear price impact function and
empirical square root impact function are considered in the price
dynamics, and power-law behaviors are observed for the return
variance $\sigma^2$, the price impact $H$ and the predictability $K$
at the large values of the parameter $N$. The sum of the traders'
wealth displays a positive value for the trading model with square
root price impact function. An unsymmetrically distribution of the
conditional excess demand $\langle A|u \rangle$ is observed, and
based on this observation we give a qualitative explanation for the
positive wealth sum for the model with square root price impact
function. In addition, an evolution mechanism is introduced to the
trading model. The introduction of new traders with randomly
selected strategies breaks down the domination of the strategies
highly performed in the model without evolution, and leads to a
relatively small value for the biased tendency of the price change,
as well as $\sigma^2$, $H$ and $K$. Power-law behavior is observed
for the time $t'$ at which $P_s$ percentage of the traders are
washed out, and the relative wealth difference between the rich and
poor traders becomes much smaller when the time approaches the point
that all the traders are washed out. The traders with the MG's
strategies are also introduced to the trading mode. The small
friction of the mixed traders are fed by the rest majority of
traders, and thus have relatively more wealth. We also introduce the
traders known as producers to this trading model, and find it leads
to an effectively increase of the traders' wealth.

\bigskip

{\textbf{Acknowledgments:}}

During this work, we have enjoyed the support and hospitality of
ISI, Torino, Italy and Renmin Univ, Beijing, China.

\bibliography{E:/Paper/bibfile/BibliographyRen}

\end{document}